\documentclass[a4paper,12pt]{article}
\usepackage{graphicx}
\usepackage{dcolumn}
\usepackage{bm}
\usepackage{amssymb}
\usepackage{amsmath}
\usepackage{hyperref}
\usepackage{authblk}
\usepackage{float}
\usepackage{caption}

\usepackage[symbol*]{footmisc}
\DefineFNsymbolsTM{myfnsymbols}{
  \textdagger    \dagger
  \textasteriskcentered *  
  \textsection   \mathsection  
  \textbardbl    \|%
  \textdaggerdbl \ddagger
  \textparagraph \mathparagraph
}%
\setfnsymbol{myfnsymbols}

\title{Particle states of Lattice QCD}
\author[1]{A.~S.~Kapoyannis\footnote{email: akapog@phys.uoa.gr}}
\author[2]{A.~D.~Panagiotou\footnote{email: apanagio@phys.uoa.gr}}
\affil[1,2]{Nuclear and Particle Physics Section, Faculty of Physics, University of Athens, GR-15784 Greece}
\date{}

\begin{document}

\maketitle

\begin{abstract}
We determine the degeneracy factor and the average particle mass of particles that produce the Lattice QCD 
pressure and specific entropy at zero baryon chemical potential.
The number of states of the gluons and the quarks are found to converge above $T=$230 MeV to almost constant
values, close to the number of states of an ideal Quark-Gluon Phase, while their assigned masses retain high
values.
The number of states and the average mass of
a system containing quarks in interaction with gluons
are found to
decrease steeply with increase of temperature between $T \sim 150$ and 160 MeV, a region contained within the
region of the chiral transition.
The minimum value of the number of states within this temperature interval indicates that the states are of
hadronic nature.
\end{abstract}



\section{Introduction}

The results from heavy ion collisions at RHIC and LHC suggest that the quark-gluon state formed is a strongly 
interacting one. The fact that the
LHC at the highest energy of $\sqrt{s}_{NN}$= 5.02 TeV produced collision temperatures of more than 300 MeV, 
pushed the ``ideal state'' of non-interacting entities to enormous temperatures.

Hagedorn, a pioneer in the investigation of the hadronic properties, had suggested that a strongly interacting 
system can be equivalently described as a system of non-interacting entities with corresponding masses. In this 
picture the interaction is hidden in ``particle'' states
\cite{Hagedorn}.
Recently, a quasi-particle using massive gluon states for the thermodynamic description of non-abelian 
interacting gluon plasma has been developed \cite{quasi-particle}.

Given the above, we may reasonably ask if it is possible to find ``particle'' massive states, which could 
equivalently describe the existing, precise results of lattice calculations for the equation of state of QCD 
matter.
 
In this work we study this possibility. In section 2 we review the Lattice calculations which are the input
of our calculations.
In section 3 we describe the method we will use and apply it to a simple, instructive model.
In section 4 we present our results firstly on the Hadron Resonance Gas (HRG) and then on the QCD system. 
Specifically, we fit the lattice
pressure and specific entropy with a model of two parameters, the number of ``degenerate'' states and the average 
mass and explore the
variation of these two parameters with temperature. Finally, we discuss our conclusions in section 5.

\section{Lattice calculations}

Lattice calculations have been carried out for (2+1) flavor QCD, i.e. for two light and one strange quark state at
vanishing baryon chemical potential \cite{2+1flavor}. In Figure 2 of \cite{2+1flavor}, the authors present their 
estimation of suitably normalised pressure $P$, entropy density $s$ and energy density $\epsilon$, of a QCD system 
in the temperature range 130-400 MeV. The transition temperature is taken to be $T_c \simeq 154$ MeV, while the 
smooth curves suggest the existence of a crossover region. In Table I of \cite{2+1flavor} the numerical values of 
the previous quantities with their corresponding errors are recorded. Also, in eq.~(16) of \cite{2+1flavor} a 
parametrisation is given which fits quite well the numerical values of the pressure curve. This parametrisation 
agrees with the lattice data well above 400 MeV. The curves approach the corresponding Hadron Resonance Gas (HRG) 
curves for temperatures $T \leq 180$ MeV. This is expected since HRG models \cite{HRG} describe well the freeze-
out data from the heavy-ion collisions.

\begin{figure*}[h]
\centering
\includegraphics[scale=0.59,angle=0]{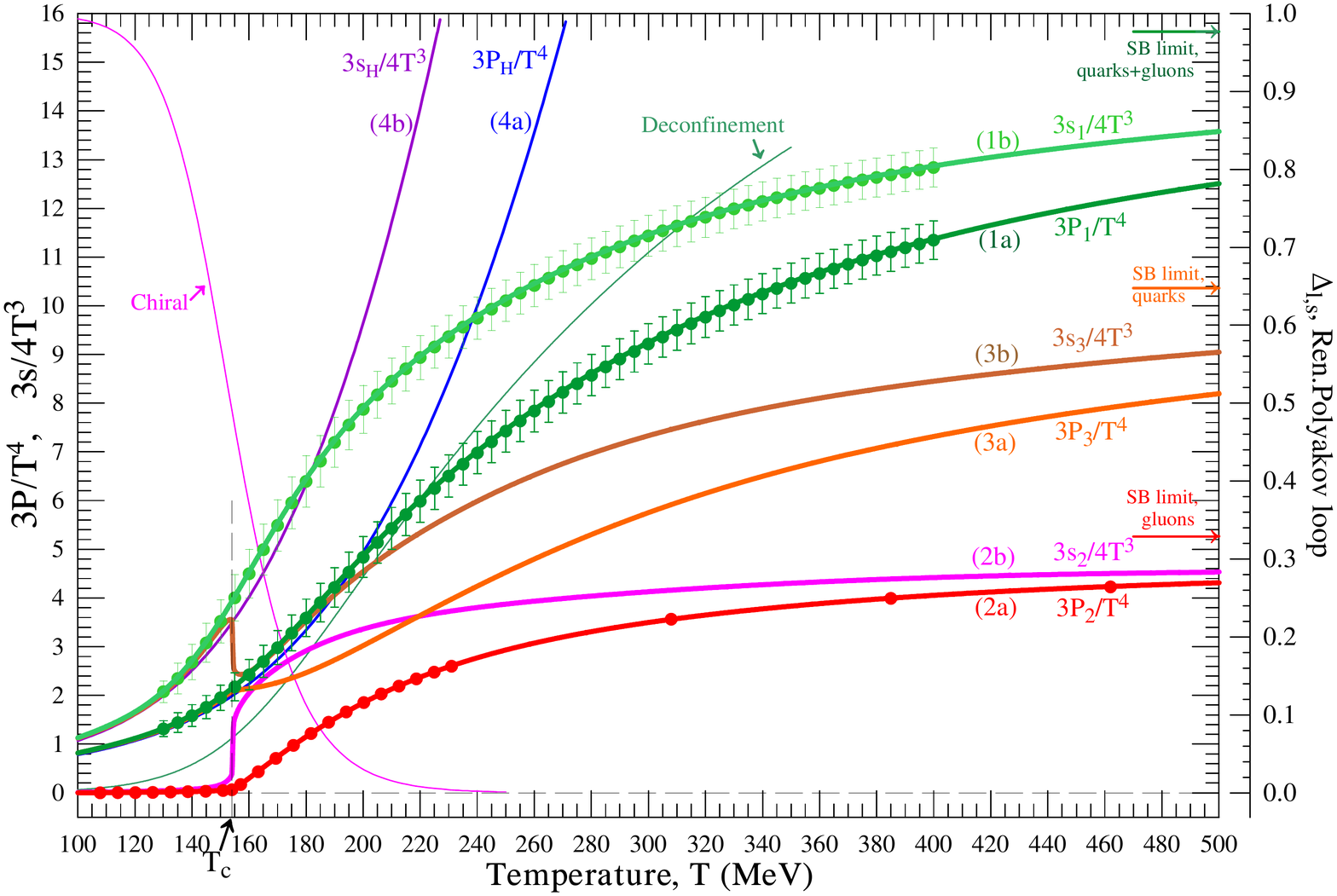}
\caption{\label{fig:P,s} {\small Calculation of the normalised Pressure, $P$ (curves a) and entropy density, $s$ 
(curves b). Lattice calculation for (2+1) flavour  QCD correspond to lines 1, lattice calculations for the SU(3) 
gauge field to lines 2 and lines 3 to the result of 
subtracting the values of lines 2 from the corresponding values of lines 1.
Estimation for the Hadron Resonance Gas (HRG) 
corresponds to lines 4. Also shown are the corresponding non-interacting (SB) limits and the lattice estimation of 
the order parameters of the chiral transition and deconfinement (corresponding to the far right vertical axis).}}
\end{figure*}

The system described by the full QCD Lagrangian including the fermionic and gluonic fields and all 
the interactions among them will be referred in
this paper as system 1. We will use the corresponding Lattice calculations for this system, which we depict in 
curves 1 of Figure \ref{fig:P,s}.
In curve (1a) of Figure \ref{fig:P,s} we depict the numerical calculations for the normalised pressure 
$\frac{3P_{1}}{T^4}$, as well as the parameterisation of \cite{2+1flavor}. In curve (1b) the numerical 
calculations for the normalised specific entropy $\frac{3s_{1}}{4T^3}$ are shown. Assuming that the pressure $P$ 
does not depend on the system volume, the corresponding curve is found from the parametrisation of curve (1a), 
since
\begin{equation} \label{eq:entropy}
s=\frac{dP}{dT}
\end{equation}
In Figure \ref{fig:P,s} we also show our calculations of the curves of normalised pressure and entropy of the 
Hadron Resonance Gas (HRG) model (curves (4a) and (4b), respectively). The calculations have been carried out in 
the Boltzmann approximation, since in this temperature range the effect of the inclusion of the correct statistics 
is negligible at these conditions.

Lattice calculations for the gauge field of QCD have been carried out in \cite{SU(3)}, as well. In Table 1 of this 
work the pressure as a function of temperature is recorded for a wide range of values 
($0.7 \leq \frac{T}{T_c} \leq 1000$). 
The system described by the Lagrangian including only the gluonic fields and the interactions only 
among them will be referred in
this paper as system 2. We will use the corresponding Lattice calculations for this system, which we depict in 
curves 2 of Figure \ref{fig:P,s}.
We apply again $T_c=154$ MeV and record these values as solid circles in 
curve (2a) of Figure \ref{fig:P,s}, which represents the relative normalised pressure $\frac{3P_2}{T^4}$. In 
order to be able to work with these values we need a continuous and derivable function. Thus, we apply a fit on 
the values. Since
we cannot find a suitable function for the whole temperature range, we use for $T \leq T_c$ the function
\begin{equation} \label{eq:fitsu3down}
P_2(T)=a_1+a_2 T+a_3 \ln(a_4+a_5 T)
\end{equation}
For $T \geq T_c$ we find as a suitable function to fit the data the following:
\begin{equation} \label{eq:fitsu3up}
P_2(T)=b_1 \left\{1-exp\left[-b_2 (x-b_3)^{b_4}\right]\right\}+
b_5 \left\{1-exp\left[-b_6 (T-T_c)^{b_7}\right]\right\}
\end{equation}
The phase transition for the pure SU(3) gauge system is of first order. This results in discontinuity
in the first derivative of pressure and in the entropy density.
This discontinuity can be problematic in method 1 we will describe in the next section, since
the method cannot be applied around the critical temperature. 
In order to avoid this discontinuity between the functions (\ref{eq:fitsu3down}) and 
(\ref{eq:fitsu3up}), we interpolate with a polynomial of 
5$\rm ^{th}$ grade in the temperature range $T_c \leq T \leq T_c +0.5 MeV$:
\begin{equation} \label{eq:fitsu3middle}
P_{2}(T)=c_1+c_2 T+c_3 T^2+c_4 T^3+c_5 T^4+c_6 T^5
\end{equation}
However, we have to note that our calculations in this small temperature range where we impose the interpolation will have to
be taken cautiously into account in the sense they may produce continuities where discontinuities exist.
We impose the constraint that the polynomial will produce continuity up to 2$\rm ^{nd}$ derivative at the boundary 
values. The values of the parameters of our fitting functions are listed in Table \ref{tab:parameters}.
We prefer to use eqs.~(\ref{eq:fitsu3down})-(\ref{eq:fitsu3middle}) instead of the functions listed in \cite{SU(3)} because the latter are valid only at certain temperature regions, while we need a function valid at
the whole temperature region on which we work.

Line (2b) in Figure \ref{fig:P,s} represents the corresponding normalised entropy density, 
$\frac{3 s_2}{4 T^3}$, as it is calculated from eq.~(\ref{eq:entropy}), using 
eqs.~(\ref{eq:fitsu3down})-(\ref{eq:fitsu3middle}).

\begin{table}[h]
\centering
\begin{tabular}{|c||c||c|} \hline
Symbol/Value        &Symbol/Value    & Symbol/Value        \\ \hline
$a_1=$0.06398	    & $b_1=$0.81546  & $c_1=$-1125870000   \\
$a_2=$0.00001	    & $b_2=$0.00279  & $c_2=$36430317      \\
$a_3=$-0.00692	    & $b_3=$147.3457 & $c_3=$-471516.11    \\
$a_4=$32351.103  	& $b_4=$1.15198  & $c_4=$3051.3982     \\
$a_5=$-206.61058	& $b_5=$0.69596  & $c_5=$-9.8734740    \\
                    & $b_6=$0.01106  & $c_6=$0.012779088   \\
                    & $b_7=$1.15973  &                     \\ \hline
\end{tabular}
\caption{\label{tab:parameters} {\small The parameters used in the fitting functions of the SU(3) pressure.}}
\end{table}

We will also use as input in our calculations the pressure and entropy density which result from 
subtracting the corresponding quantities of system 2 from those of system 1:
\begin{equation} \label{eq:P3,s3}
P_3=P_1-P_2\;\;,\;s_3=s_1-s_2\;.
\end{equation}
Equations (\ref{eq:P3,s3}) describe a system of fermionic fields in interaction with gluons which will be referred to in this work as system 3.
This system differs from the total system 1 in the exclusion of the contribution of the gluons (system 2).
This separation cannot in principle be made in an interacting theory.
However, this separation is useful in the sense that, as will  become evident in section 4, at temperatures $T>230$ MeV system 3 leads to a system containing only quark degrees 
of freedom, while system 2 to a system with only gluonic degrees.
Also, at temperatures below the critical one ($T<154$ MeV) system 3 converges to system 1, acquiring the hadronic degrees of freedom,
while system 1 results to negligible contribution to pressure and entropy density (Figure \ref{fig:P,s}).
The curves corresponding to system 3 are curve (3a) for the normalised quark pressure and curve (3b) for the 
normalised entropy density, $\frac{3 s_3}{4 T^3}$.

In Figure \ref{fig:P,s} we depict with horizontal lines the Stefan-Boltzmann (SB) limit for massless and 
non-interacting particles. For gluons (bosons) the pressure is given by
\begin{equation} \label{eq:PgSB}
\frac{P_g}{T^4}=\frac{\pi^2}{90} g_G \equiv \frac{\pi^2}{90} g_s g_c =\frac{\pi^2}{90} 2 \cdot 8 =\frac{\pi^2}{90} 16 \;,
\end{equation}
where $g_G=16$ is the total number of states of gluons due to their spin, $g_s=2$ and colour, $g_c=8$.
For the quarks (fermions) the pressure is given by
\begin{equation} \label{eq:PqSB}
\frac{P_q}{T^4}=\frac{7}{8} \frac{\pi^2}{90} g_Q \equiv \frac{7}{8}\frac{\pi^2}{90} g_s g_c g_f g_{an}=\frac{7}{8}\frac{\pi^2}{90} 2 \cdot 3 \cdot 3 \cdot 2 =\frac{7}{8}\frac{\pi^2}{90} 36\;,
\end{equation}
where $g_Q=36$ is the total number of states of quarks due to their spin, $g_s=2$, colour, $g_c=3$, flavours which 
are present in (2+1) flavour QCD, $g_f=3$ and the presence of quarks and anti-quarks, $g_{an}=2$.
Normalising the entropy density as $\frac{3s}{4T^3}$ we get the same SB limit as the pressure $\frac{3P}{T^4}$.

In \cite{ch-d} (Figure 4, Table 3) there is calculation of $\Delta_{l,s}$ as a function of temperature, a quantity related to the chiral condensate, which can be taken as an indicator for the remnant of the chiral transition, since it shows abrupt change.
Also, in \cite{ch-d} (Figure 3, Table 3) there is an estimation of the renormalised Polyakov loop, the order parameter of the deconfinement phase transition of QCD in the pure gauge sector. We depict in Figure \ref{fig:P,s} our fits to the values of these two quantities listed in \cite{ch-d}.

\section{Method}

We address the question which single particle state would have the same pressure with the lattice QCD calculations 
at a particular temperature. For this reason we assume that the partition function of this particle state at 
vanishing baryon chemical potential will be
\begin{equation} \label{eq:lnZfit_Bo}
\ln Z_f (V,T; g_t,\left<m\right>) = \frac{VT}{2\pi^2}g_t \left<m\right>^2 K_2\left(\frac{\left<m\right>}{T}\right) \;.
\end{equation}
We use the above partition function, which is in the Boltzmann approximation, in the absence of knowledge whether 
the particles of our system are fermions, bosons or a mixture of both.
Also, the correct statistics are not expected to produce high differences at the temperatures considered. 
The pressure and entropy density related to (\ref{eq:lnZfit_Bo}) will then be
\begin{equation} \label{eq:Pfit}
P_f (T;g_t,\left<m\right>) = \frac{T^2}{2\pi^2}g_t \left<m\right>^2 K_2\left(\frac{\left<m\right>}{T}\right) \;,
\end{equation}
\begin{equation} \label{eq:sfit}
s_f (T;g_t,\left<m\right>) = \frac{g_t \left<m\right>^2}{2\pi^2}
\left \{ 2T K_2\left(\frac{\left<m\right>}{T}\right)+
\frac{\left<m\right>}{2} 
\left[K_1\left(\frac{\left<m\right>}{T}\right)+
K_3\left(\frac{\left<m\right>}{T}\right)\right]\right\}\;.
\end{equation}
The problem is, henceforth, limited to determining the quantity $\left<m\right>$, which corresponds to the mass of a 
particle, as well as the degeneracy factor $g_t$, which is the total number of states that corresponds to this 
particle (due to its spin, colour, etc.). To address the problem we have developed two methods.

In method ``1'' we consider a part of the Lattice pressure curve $P_L(T)$ which extends in the temperature 
interval $(T-\Delta T,T+\Delta T)$ around a central value $T$. We divide the interval in $N-1$ equal parts, 
with $N$ being rather large (we use for all calculations in this paper $N=101$).
The corresponding temperature values are $T_i$, $i=1,N$. 
Then we determine $g_t$ and $\left<m\right>$ by minimizing the quantity
\begin{equation} \label{eq:chi2}
\chi^2=\sum_{i=1}^{N}\frac{1}{\sigma_i^2}\left[P_L(T_i)-P_f(T_i;g_t,\left<m\right>)\right]^2 \;,
\end{equation}
The determined values $g_t$ and $\left<m\right>$ can then be assigned to the central temperature 
value $T$.
In the definition of $\chi^2$, $\sigma_i$ are errors in the values $P_L(T_i)$.
Since all the points $i$ should have the same weight in the
calculation these errors should be equal for the same central temperature $T$. However, to compare values of
$\chi^2$ at different temperatures $T$ we should consider the fact that pressure rises rapidly with $T$.
Consequently, it is expected that the deviations between the pressure $P_L$ and the fitting pressure $P_f$ to 
increase with the rise of $T$, if we use constant values for $\sigma_i$ for all $T$.
In order to compensate this fact and avoid the systematic growth of $\chi^2$ with temperature we use for the
errors the function 
\begin{equation} \label{eq:sigma}
\sigma (T) =f(T)=T^4 \;,
\end{equation}
which has also been used extensively for the normalisation of pressure. We note, though, that the exact position
of the maxima of $\chi^2$ will depend on the function used in (\ref{eq:sigma}), while this choice will cause
no effect on the determination of the parameters $g_t$ and $\left<m\right>$, as it is evident from 
eq.~(\ref{eq:chi2}).

In method ``2'' we use the values of the Lattice pressure $P_L(T)$ and the Lattice energy density $s_L(T)$ at a 
specific temperature $T$. Then we solve the following set of two equations to determine $g_t$ and
$\left<m\right>$ which correspond to the specific value of $T$:
\begin{equation} \label{eq:P,s}
P_L(T)=P_f(T;g_t,\left<m\right>)\;\;\;,\;\;\;s_L(T)=s_f(T;g_t,\left<m\right>)
\end{equation}

The choice of the two thermodynamic functions $P_L(T)$ and $s_L(T)$, which we use as input in method 2, is not unique.
Let us suppose, on the event of use of the pair $P_L(T)$ and $s_L(T)$, that we want to fit another thermodynamic quantity
$r_L$ which can be expressed as function of the used two quantities,
\begin{equation} \label{eq:r,P,s}
r_L(T)=F(P_L(T),s_L(T))\;.
\end{equation}
We want to fit the quantity $r_L$ with the same parameters $g_t$ and $\left<m\right>$. So
\begin{equation} \label{eq:r_Lf}
r_L(T)=r_f(T;g_t,\left<m\right>)\;,
\end{equation}
where $r_f$ is the fitting function which corresponds to the quantity $r$ and which is calculated from the partition function 
for a system with one particle species (like eqs.~(\ref{eq:Pfit}),(\ref{eq:sfit})).
Combining eqs.~(\ref{eq:r,P,s}), (\ref{eq:r_Lf}) and (\ref{eq:P,s}), we find that the following equation has to hold for every
temperature $T$:
\begin{equation} \label{eq:r,F}
r_f(T;g_t,\left<m\right>)=F(P_f(T;g_t,\left<m\right>),s_f(T;g_t,\left<m\right>))\;.
\end{equation}
Now if $g_{ta}$, $\left<m\right>_a$ is the solution of eqs.~(\ref{eq:P,s}) for the parameters for a specific temperature
$T_a$, the quantity $r_L$ will be at the same temperature:
\begin{equation} \label{eq:r,proof}
r_{La} \equiv r_L(T_a) = F(P_L(T_a),s_L(T_a)) =
F(P_f(T;g_{ta},\left<m\right>_a),s_f(T;g_{ta},\left<m\right>_a)) =
r_f(T;g_{ta},\left<m\right>_a)
\end{equation}
The last equation verifies that the parameters $g_t$ and $\left<m\right>$ that were found to fit the pressure and entropy 
density at a certain temperature will also fit exactly the quantity $r_L$.
If the partition function does not depend on the system volume and if we work at vanishing chemical potential,
we can use as function $r_L$ the entropy density $\varepsilon_L$, since:
\begin{equation} \label{eq:e_L}
\varepsilon_L(T)=-P_L(T)+T s_L(T),
\end{equation}
or the trace anomaly, used extensively by lattice practitioners, since
\begin{equation} \label{eq:Th_L}
\Theta^{\mu\mu}_L(T)=\varepsilon(T)-3P_L(T)=T s_L(T)-4 P_L(T).
\end{equation}
One can equivalently use in method 2 any two of a set of thermodynamic quantities that can be expressed as a function of each other
to determine the two parameters $g_t$ and $\left<m\right>$.

\begin{figure*}[h]
\centering
(i)\includegraphics[scale=0.27,angle=0]{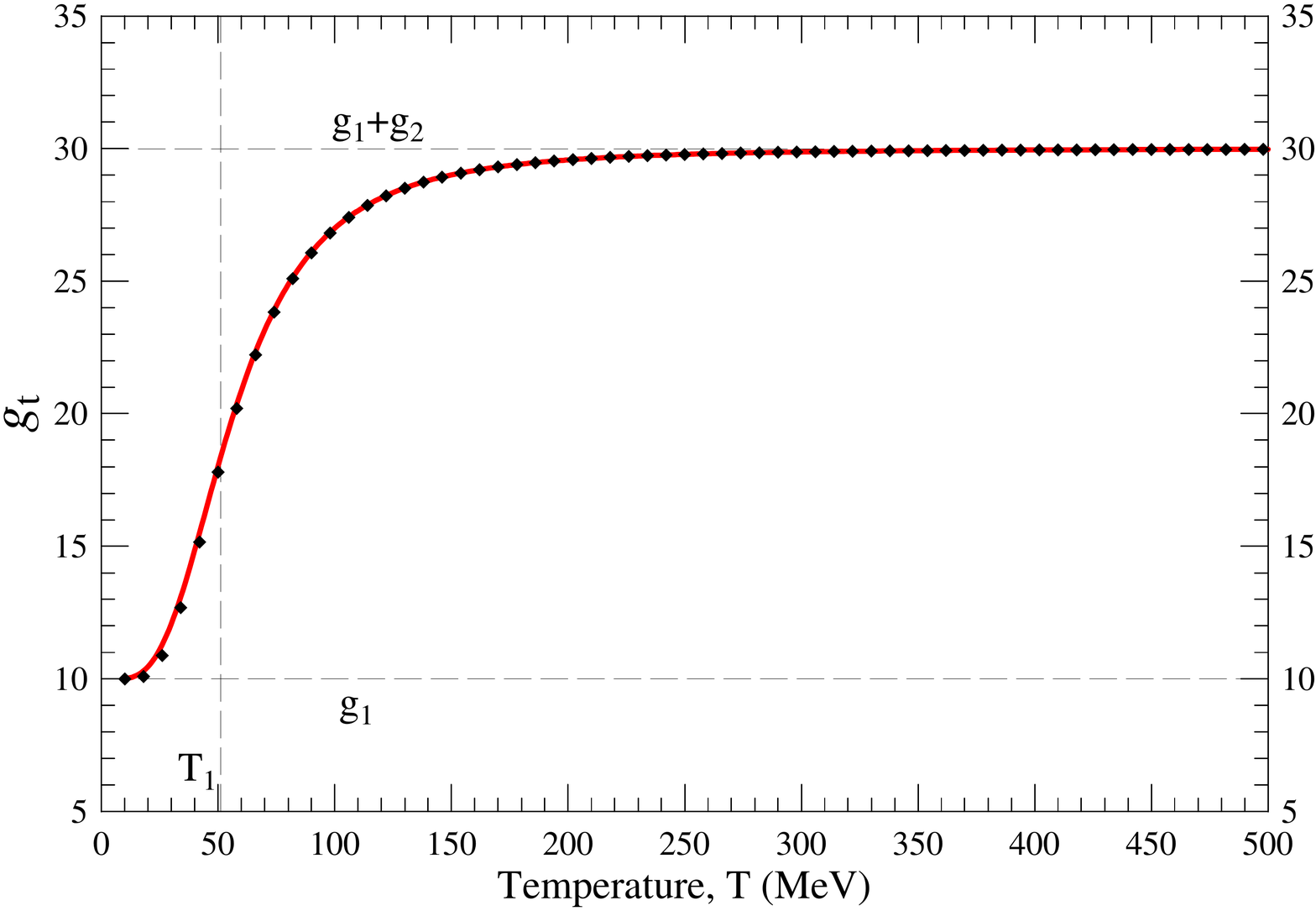}
(ii)\includegraphics[scale=0.27,angle=0]{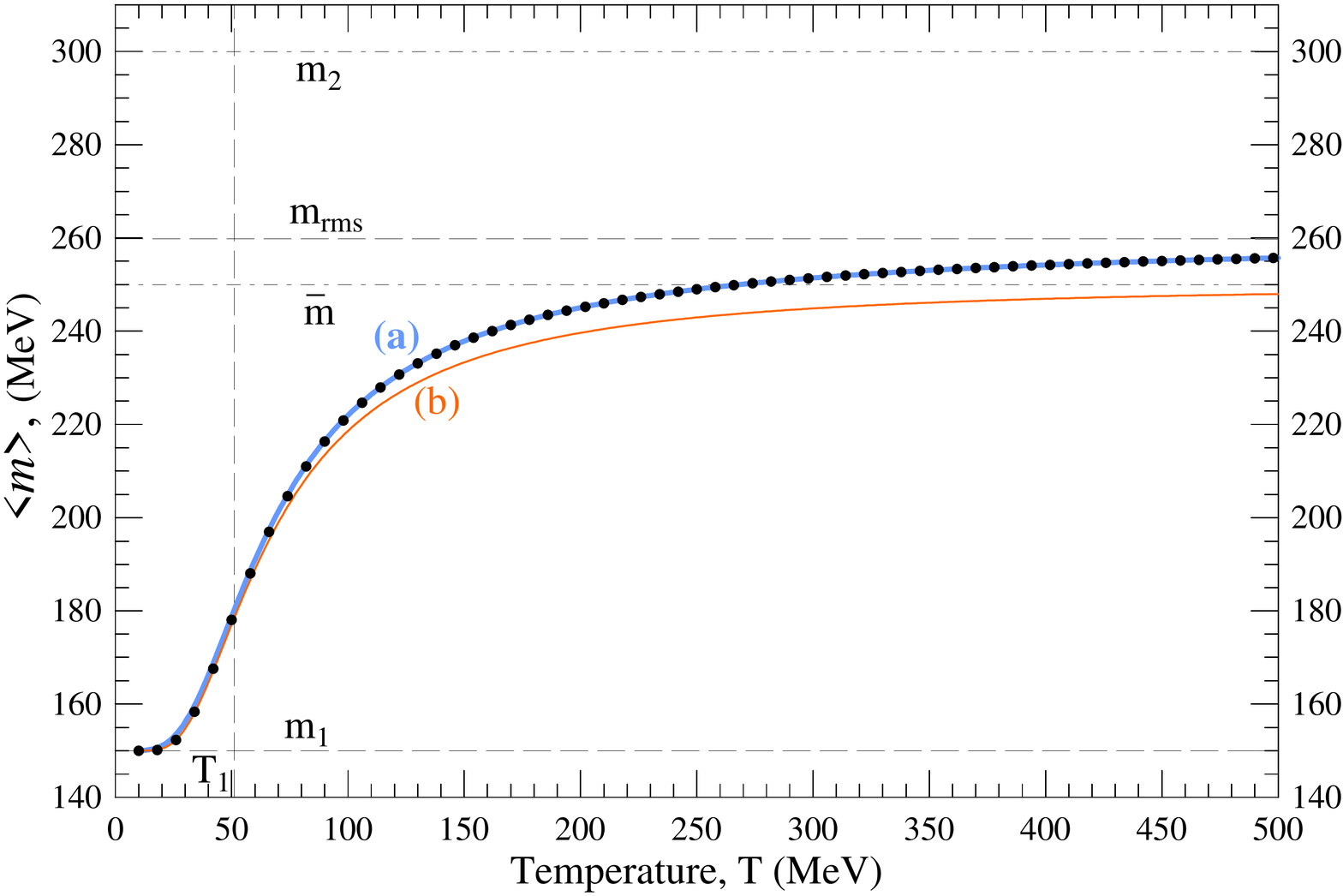}
(iii)\includegraphics[scale=0.27,angle=0]{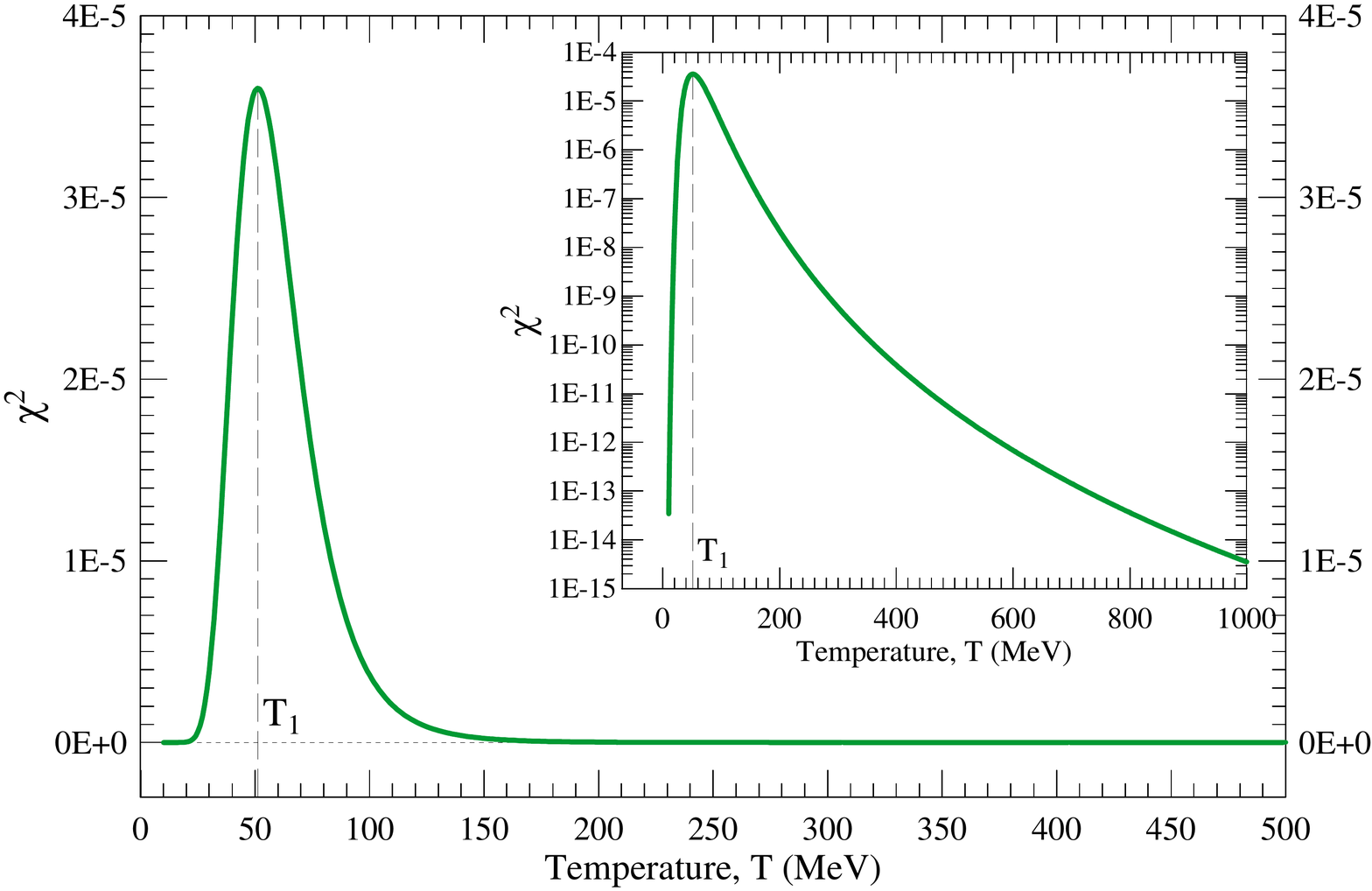}
\caption{\label{fig:gmx2} {\small Analysis of the partition function of a system with two particle species with
degeneracy factors $g_1=10$ and $g_2=20$ and mass $m_1=150$ MeV and $m_2=300$ MeV, respectively. 
(i) $g_t$ as a function of temperature with method 1 (continuous line) and method 2 (solid squares). 
(ii) $\left<m\right>$ as a function of temperature (curve (a)) with method 1 (continuous line) and method 2 
(solid circles). Curve (b) represents an approximation valid for low temperatures.
(iii) The value of $\chi^2$ as a function of temperature for method 1.}}
\end{figure*}

Comparing the results from the two methods it is deduced that they are in general equivalent.
This is expected since in method 1 we fit the shape of a
segment
of the pressure curve. By taking a lot of points on this part and requiring these points to be fitted by 
our theoretical pressure function we impose the constraint not only to fit the specific value of $P_L(T)$ at $T$ 
but also the higher derivatives of the function $P_L(T)$ at $T$. In method 2, according to eqs.~(\ref{eq:P,s}) we 
fix our theoretical functions to fit the Lattice pressure and its first derivative (entropy density). Thus, we 
expect the outcome from the two methods to be approximately the same. 

However, method 1 gives us the additional 
opportunity to calculate the value of $\chi^2$. This value can be taken as a measure of how well function 
(\ref{eq:Pfit}) fits the Lattice pressure. On the other hand, in method 1 
the results may depend on the value of the temperature interval $\Delta T$. These issues will be discussed in the 
following section.

Additionally, since the pressure can be expressed as an integral of the entropy density we could start by fitting in method 1 
a segment of the curve of the entropy density. Thus, the choice of the pressure in method 1 is not unique.

\vspace{-0.0cm}
We shall end this section by performing some ``toy'' fits that will enable us to have insight on the meaning 
of the parameters $g_t$ and $\left<m\right>$. In the first case we consider a system composed of one boltzmannian 
particle with degeneracy factor $g_1$ and mass $m_1$. The relevant pressure is then 
\begin{equation} \label{eq:P1}
P_a (T) = \frac{T^2}{2\pi^2}g_1 m_1^2 K_2\left(\frac{m_1}{T}\right) \;.
\end{equation}
If we perform method 1 or method 2 we find $g_t=g_1$ and $\left<m\right>=m_1$. So in this case $g_t$ can be 
identified as the number of the states of the particle and $\left<m\right>$
as the mass of this particle.

We then consider a system composed of two particles with pressure
\begin{equation} \label{eq:P2}
P_b (T) = \frac{T^2}{2\pi^2} g_1 m_1^2 K_2\left(\frac{m_1}{T}\right) +
\frac{T^2}{2\pi^2} g_2 m_2^2 K_2\left(\frac{m_2}{T}\right)\;.
\end{equation}

In figures \ref{fig:gmx2} we present the results for $g_t$ (fig.~(i)) and $\left<m\right>$ (fig.~(ii)) as a 
function of temperature. 
We will determine the low and high temperature limits of these two entities. The modified Bessel functions of the 
2nd kind can be approximated for large arguments (low temperatures) as
\begin{equation} \label{eq:approxK-lowT}
K_2 (z) \underset{z \rightarrow \infty}{\longrightarrow} \sqrt{\frac{\pi}{2z}} e^{-z} \Rightarrow 
K_2\left(\frac{m}{T}\right) \underset{T \rightarrow 0}{\longrightarrow} \sqrt{\frac{\pi}{2}\frac{T}{m}} e^{-m/T} \;.
\end{equation}
From this we get
\[
P_b (T) = P_f(T;g_t,\left<m\right>) \underset{T \rightarrow 0}{\Rightarrow}
g_1 m_1^2\sqrt{\frac{\pi}{2}\frac{T}{m_1}} e^{-m_1/T}+
g_2 m_2^2\sqrt{\frac{\pi}{2}\frac{T}{m_2}} e^{-m_2/T}=
g_t \left<m\right>^2\sqrt{\frac{\pi}{2}\frac{T}{\left<m\right>}} e^{-\left<m\right>/T} \Rightarrow
\]
\begin{equation} \label{eq:approxeq-lowT}
g_1 m_1^{3/2} e^{(\left<m\right>-m_1)/T}+g_2 m_2^{3/2} e^{(\left<m\right>-m_2)/T}=
g_t \left<m\right>^{3/2}
\end{equation}
For eq.~(\ref{eq:approxeq-lowT}) to be valid, the exponential with the greater exponent has to be eliminated. 
Since $m_1<m_2$, this requirement leads to 
\begin{equation} \label{eq:approxm-lowT}
\left<m\right> \underset{T \rightarrow 0}{\longrightarrow} m_1
\end{equation}
Automatically, the other exponential acquires negative exponent and tends to zero. Thus
\begin{equation} \label{eq:approxg-lowT}
g_t \underset{T \rightarrow 0}{\longrightarrow} g_1
\end{equation}
Eqs.~(\ref{eq:approxm-lowT}),(\ref{eq:approxg-lowT}) prove what is evident from figures \ref{fig:gmx2}(i)-(ii). 
For low temperatures the system can effectively be described as a system containing only states with the lowest 
mass $m_1$. However, as the temperature rises, the states with mass $m_2$ begin to ``unlock'', i.e.
affecting the system.
To find the effective description of our system for large temperatures we need an approximation of the 
$m^2 K_2 (m/T)$ function for small arguments that will depend on $m$.
Thus we use the following approximation of the modified Bessel functions of the 2nd kind for small arguments
\begin{equation} \label{eq:approxK_highT}
K_2(z) =\frac{z}{4}[K_3(z)-K_1(z)] \underset{z \rightarrow 0}{\longrightarrow}
 2z^{-2}-\frac{1}{4} \Rightarrow
m^2 K_2\left(\frac{m}{T}\right) \underset{T \rightarrow \infty}{\longrightarrow} 2 T^2 -\frac{m^2}{4}\;.
\end{equation}
Last equation leads to
\[
P_b (T) = P_f(T;g_t,\left<m\right>) 
\underset{T \rightarrow \infty}{\Rightarrow}
g_1 \left(2T^2-\frac{m_1^2}{4}\right)+
g_2 \left(2T^2-\frac{m_2^2}{4}\right)=
g_t \left(2T^2-\frac{\left<m\right>^2}{4}\right) \Rightarrow
\]
\begin{equation} \label{eq:approxeq-highT}
2(g_1+g_2-g_t)T^2-\frac{1}{4} (g_1 m_1^2+g_2 m_2^2-g_t\left<m\right>^2)=0
\end{equation}
Demanding eq.~(\ref{eq:approxeq-highT}) to be valid, we find that $g_t$ tends to the sum of the number of states of 
all (two) particles, while the limit of $\left<m\right>$ becomes the square root of the mean value of the
square of the masses:
\begin{equation} \label{eq:approxg-highT}
g_t \underset{T \rightarrow \infty}{\longrightarrow} g_1+g_2
\end{equation}
\begin{equation} \label{eq:approxm-highT}
\left<m\right> \underset{T \rightarrow \infty}{\longrightarrow}
m_{rms}=\sqrt{\frac{g_1 m_1^2 + g_2 m_2^2}{g_1+g_2}}\;.
\end{equation}
Eq.~(\ref{eq:approxg-highT}) enables us to identify $g_t$ as the ``total'' number of particle states, while 
eq.~(\ref{eq:approxm-highT}) suggests that $\left<m\right>$ can be identified as
the ``mean'' mass of particle states. Also, another important fact from eq.~(\ref{eq:approxeq-highT}) is that
$g_t$ is connected to the leading temperature term $T^2$. 
This forces $g_t$ to approach to high temperature limit (\ref{eq:approxg-highT}) more rapidly than
$\left<m\right>$ approaches the relevant limit (\ref{eq:approxm-highT}). This attribute is evident by
comparing Figures \ref{fig:gmx2}(i)-(ii).

The mass $\left<m\right>$ can also be approximated in a wide temperature range, which are not high, by the 
average of the two masses weighted by their contribution to the pressure
(curve (b) of figure \ref{fig:gmx2}(ii)):
\begin{equation} \label{eq:approxm-mediumT}
\bar{m} = \frac{m_1 [g_1 m_1^2 K_2\left(\frac{m_1}{T}\right)] + m_2 [g_2 m_2^2 K_2\left(\frac{m_2}{T}\right)]}
{g_1 m_1^2 K_2\left(\frac{m_1}{T}\right)+g_2 m_2^2 K_2\left(\frac{m_2}{T}\right)}\;.
\end{equation}
The high temperature limit of (\ref{eq:approxm-mediumT}) with the use of only the leading term of 
(\ref{eq:approxK_highT}) is found to be:
\begin{equation} \label{eq:approxm-mediumT-highT}
\bar{m} \simeq \frac{g_1 m_1 + g_2 m_2}{g_1+g_2}\;.
\end{equation}

Figures \ref{fig:gmx2}(i)-(ii) also show the equivalence of the two methods. In Fig.~\ref{fig:gmx2}(iii) we show
the variation of $\chi^2$ as function of temperature. It is evident from the low value of $\chi^2$ that our description with eq.~(\ref{eq:lnZfit_Bo}) is valid throughout the whole temperature range. 
The value of $\chi^2$, though, can offer us additional information. In fig.~\ref{fig:gmx2} we depict with 
vertical line the position at  temperature $T_1$ of the maximum value of $\chi^2$. 
This vertical line is depicted in figures \ref{fig:gmx2}(i)-(ii), as well.
The maximisation of $\chi^2$ suggests a relatively worse description with the used parameters at $T_1$ compared with the relatively better description at lower and higher temperatures. 
At low temperatures the description is carried out by $g_t=g_1$ and 
$\left<m\right>=m_1$, while at high temperatures by $g_t=g_1+g_2$ and $\left<m\right>=m_{rms}$. Both of these
two descriptions are connected with relatively lower value of $\chi^2$.
Also, $T_1$ coincides with the temperature at which $g_t$ or $\left<m\right>$ exhibits great variation 
(with respect to the temperature).
Thus, the temperatures at which the local maximum values of $\chi^2$ are located signal transition regions from 
one effective description, with one set of values of $g_t$ and $\left<m\right>$, to another, with different set of 
values of $g_t$ and $\left<m\right>$.

\section{Results}

We will apply the methods presented in the previous section firstly to the Hadron Resonance Gas (HRG). 
The model we use does not include volume corrections, nor other interaction among the hadrons.
It is interesting to analyse HRG because it serves as the lower limit of lattice QCD. As it is evident from Figure 
\ref{fig:P,s} the HRG pressure and entropy density remains close to the corresponding lattice calculations even 
above critical temperature and up to 
about 190 MeV. Also, the HRG model we will use is a group of particle states with three flavours and masses up to 
about 2400 MeV. As such, it will enable us to verify and extend our conclusions of the two particle states toy 
model we described in the previous section. Lastly, the HRG model will be utilised to contrast the behaviour of the 
QCD system. The particles are non-interacting and, thus, exhibit no phase transition. Also, the particle masses 
remain unchanged in the entire temperature range.

\begin{figure*}[ht]
\centering
(i)\includegraphics[scale=0.27,angle=0]{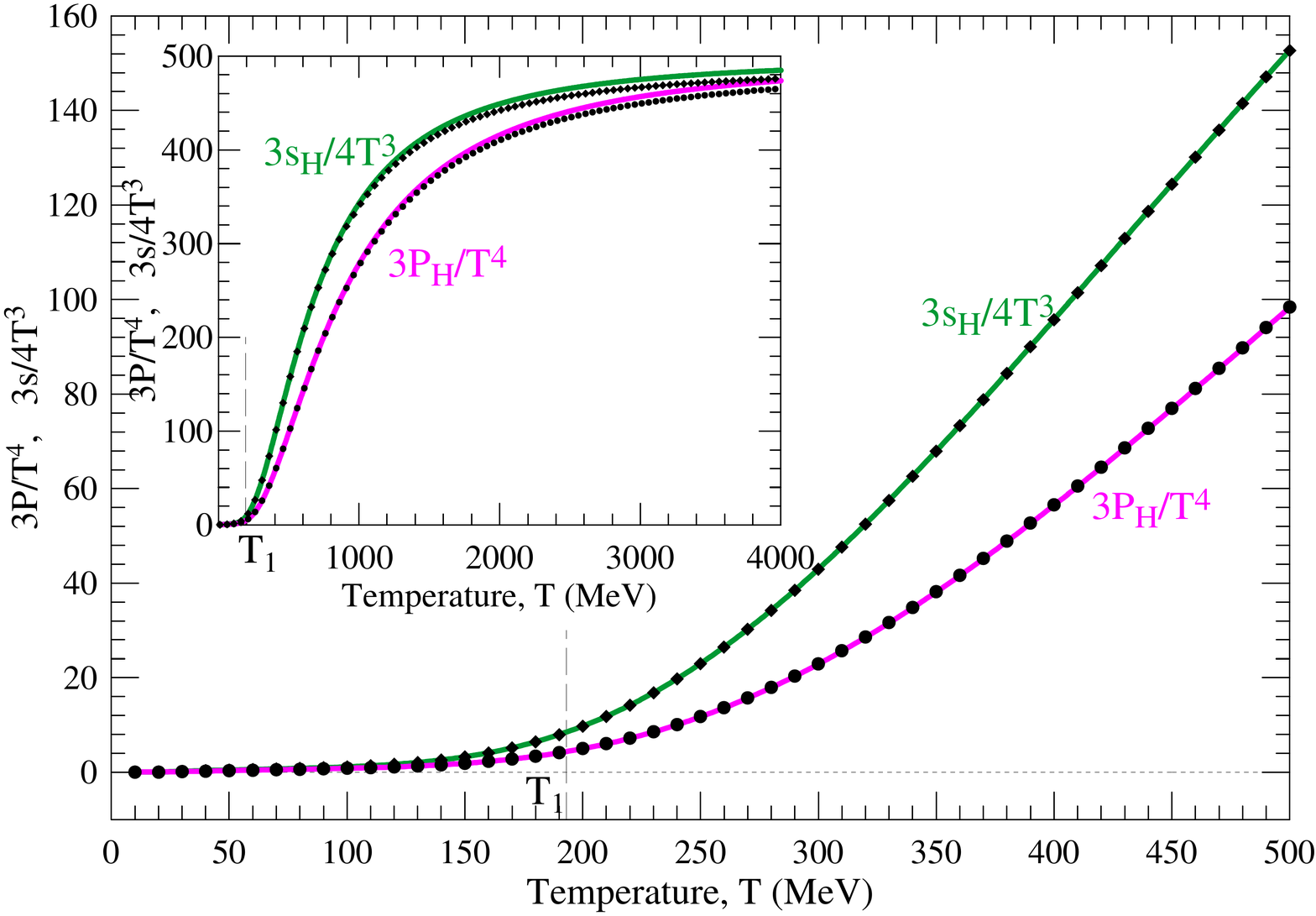}
(ii)\includegraphics[scale=0.27,angle=0]{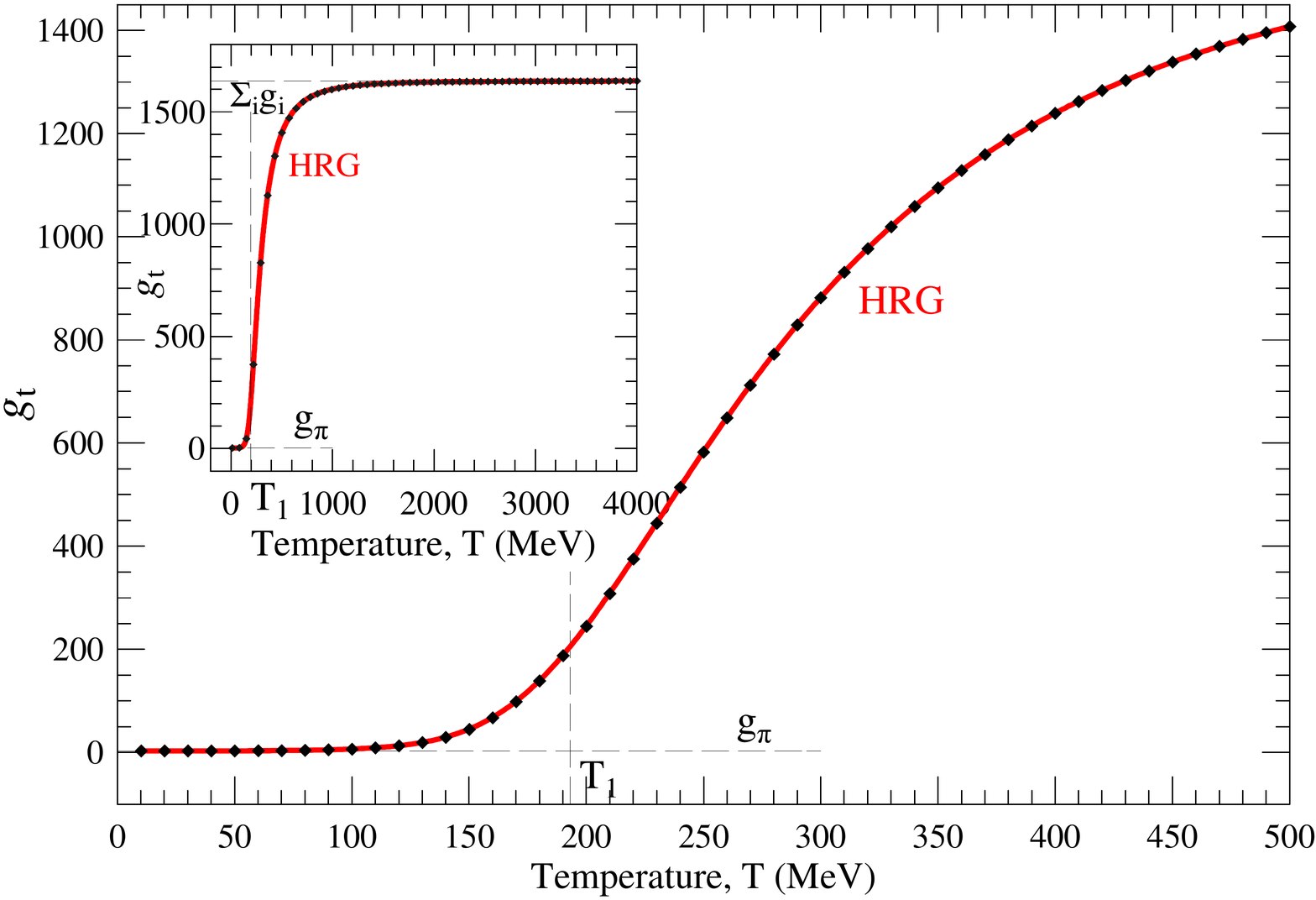}
(iii)\includegraphics[scale=0.27,angle=0]{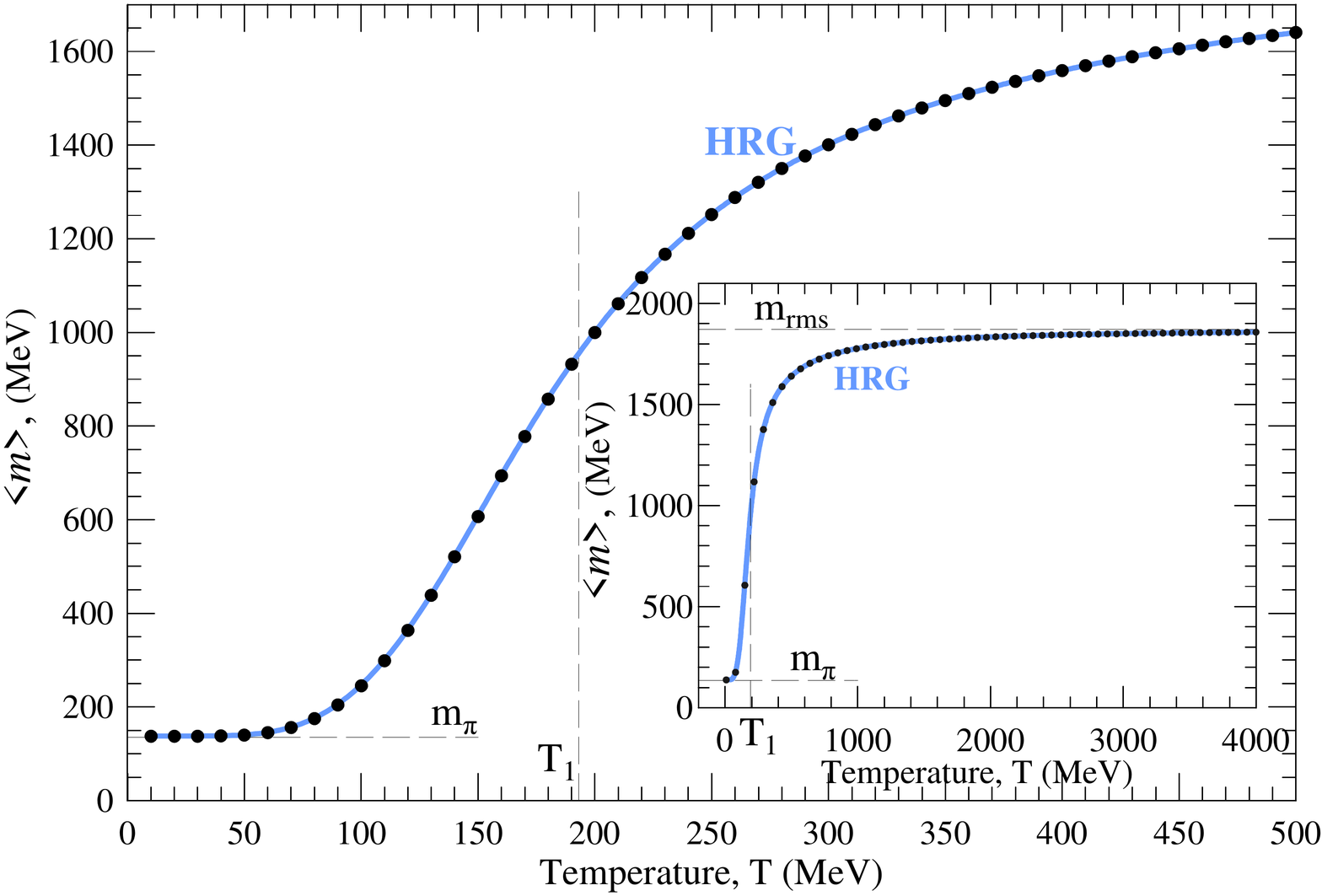}
(iv)\includegraphics[scale=0.27,angle=0]{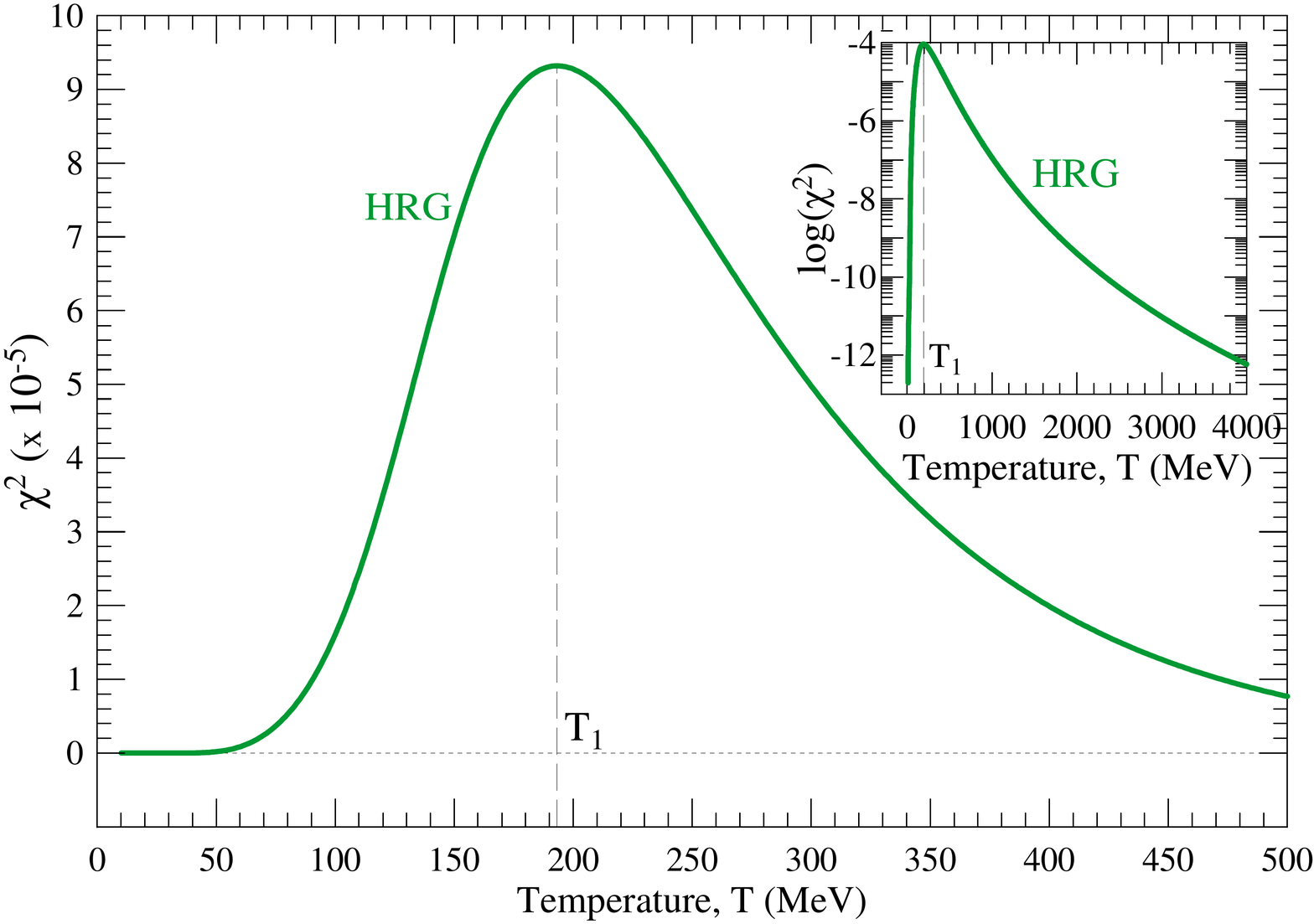}
\caption{\label{fig:HRG} {\small Analysis of the Hadron Resonance Gas (HRG). 
(i) The normalised pressure, $P_H$ and entropy density, $s_H$, of the system in the Boltzmann approximation
(lines) and with the correct statistics (Bose or Fermi) for every particle (points). The deviations between the two sets
of statistics appear in high temperatures, $T \geq 1000$ MeV, where, after all, the HRG is not valid.
(ii) $g_t$ as a function of temperature with method 1 (continuous line) and method 2 (solid squares). 
(iii) $\left<m\right>$ as a function of temperature with method 1 (continuous line) and method 2 (solid circles).
(iv) The value of $\chi^2$ as a function of temperature for method 1.}}
\end{figure*}

In Figure \ref{fig:HRG}(i) we show the pressure and the energy density of the HRG model in the Boltzmann 
approximation (continuous lines), as well as, the results with the correct statistics, Bose-Einstein or Fermi-Dirac for 
every particle (points). 
Up to $T \simeq 500$ MeV, a temperature range that will interest us in the following, the discrepancies between 
the approximation and the correct statistics are negligible.

In Figure \ref{fig:HRG}(ii) we present the calculation of the total number of states $g_t$ for the HRG as function 
of temperature with method 1 (continuous lines) and method 2 (points). The lower temperature limit of $g_t$ is 
$g_{\pi}=3$, which is the number of states of the lighter hadrons, i.e. the pions ($\pi^0$, $\pi^+$ and $\pi^-$). 
The high temperature limit of $g_t$ is checked to be the one corresponding to equation 
(\ref{eq:approxg-highT}), which for the $N$ particle species we have used is:
\begin{equation} \label{eq:gt-HRG}
g_t=\sum_{i=1}^N g_i = 1639.
\end{equation}

In Figure \ref{fig:HRG}(iii) we present the calculation of the average mass $\left<m\right>$ for the HRG as 
function of temperature with method 1 (solid lines) and method 2 (points). The lower temperature limit of 
$\left<m\right>$ is the mass of the lighter hadron, i.e. the pion. The higher temperature limit of 
$\left<m\right>$ is checked to be the corresponding to eq.~(\ref{eq:approxm-highT}), which for the $N$ particle 
species we have used is:
\begin{equation} \label{eq:m_rms-HRG}
m_{rms}=\sqrt{\frac{\sum_{i=1}^N g_i m_i^2}{\sum_{i=1}^N g_i}} \simeq 1873.65\;{\rm MeV}.
\end{equation}

In Figure \ref{fig:HRG}(iv) we show the value of $\chi^2$ of method 1. The temperature $T_1$ corresponds to the 
higher value of $\chi^2$. Comparing Figure \ref{fig:HRG}(iv) with Fig.~\ref{fig:gmx2}(iii) we find that  $\chi^2$
develops a maximum that is ``broader'' than the two particle model of the previous section. This is caused by the 
large number of resonances that exist in the HRG model, which are scattered to the whole mass range from the pion 
mass up to 2400 MeV.

\begin{figure*}[ht]
\centering
\includegraphics[scale=0.58,angle=0]{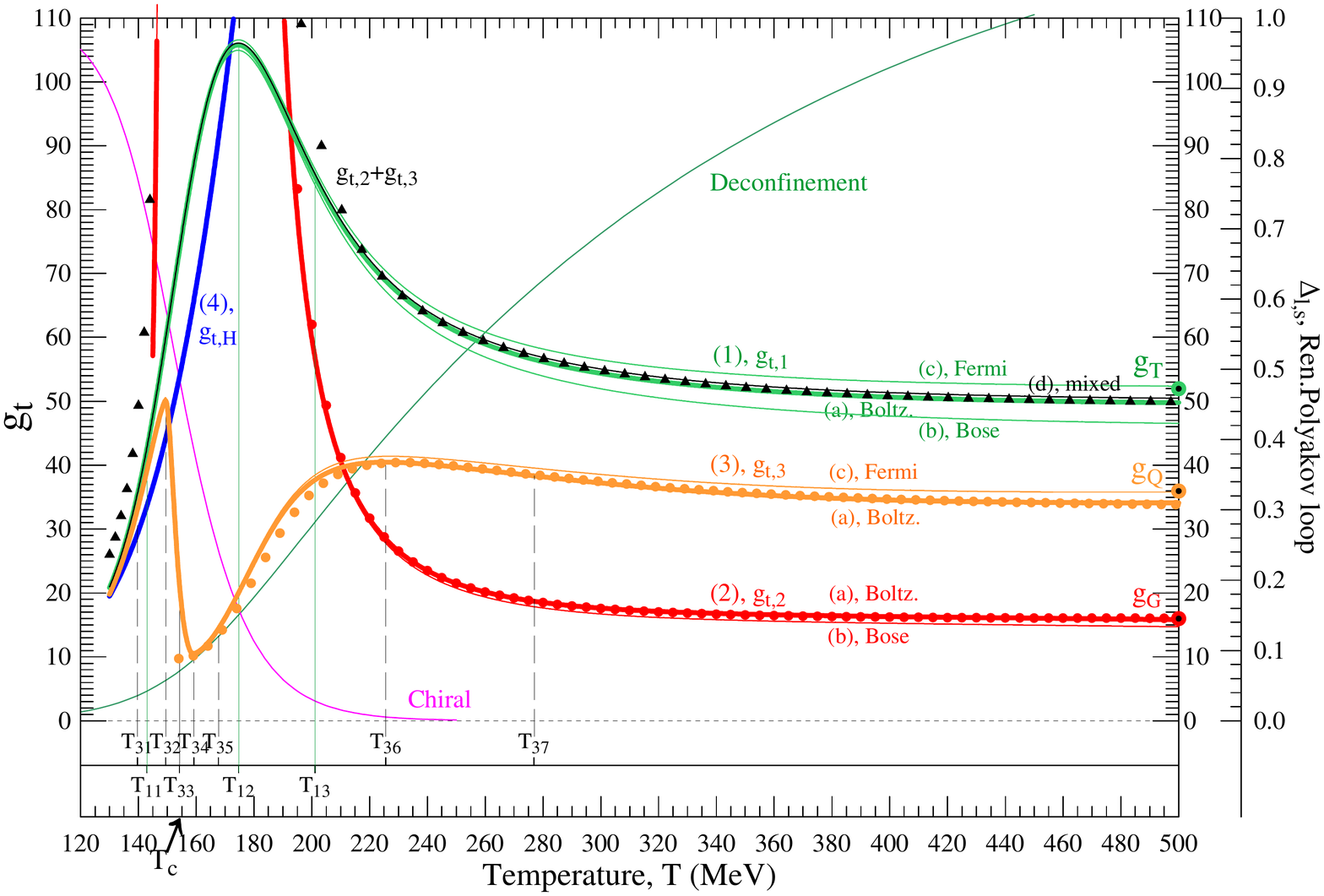}
\vspace{-0.5cm}
\caption{\label{fig:Latgt} {\small The number of states $g_t$ calculated through method 1 with $\Delta T =5$ MeV. 
Curves (1), $g_{t,1}$, correspond to system 1, curves (2), $g_{t,2}$, to system 2, curves
(3), $g_{t,3}$, to system 3
and curve (4), $g_{t,H}$, to HRG. 
Thick lines (a) are calculated with the Boltzmann approximation, lines (b) with Bose statistics, lines (c) with 
Fermi statistics and thin dark line (d) to a system composed of $\frac{16}{52}$ bosons and of $\frac{36}{52}$ 
fermions. 
The solid circles correspond to the number of states of an ideal system containing gluons 
($g_G = 16$), quarks ($g_Q = 36$) or gluons and quarks ($g_T = 52$). Solid triangles are the sum $g_{t,2}+g_{t,3}$ 
of lines (2a) and (3a) and are compared with line (1a). The order parameters of the chiral transition and deconfinement
are shown, as in Fig.~\ref{fig:P,s}. The shown characteristic temperatures $T_{31 \ldots 36}$, $T_{11 \ldots 13}$ 
are defined below in Fig.~\ref{fig:Latx2} and Table \ref{tab:x2_ex}. $T_c=154$ MeV, is the critical temperature.
With solid rectangles we present the calculations of $g_t$ in the Boltzmann approximation for systems 
2 and 3 carried out with the Lattice results from \cite{SU(3)_}.
}}
\end{figure*}

We proceed by analysing the Lattice results presented in Fig.~\ref{fig:P,s}. We limit our analysis in 
temperatures up to 500 MeV, since the calculations for (2+1) flavour QCD are limited up to 400 MeV 
(see Fig.~\ref{fig:P,s}), above which only the extrapolation function is available \cite{2+1flavor}.
In Figure~\ref{fig:Latgt} we present our calculations for the total
number of states $g_t$ and in Figure~\ref{fig:Latm} the calculations for the average mass $\left<m\right>$. These
calculations have been carried out through method 1. 
Curves (1) correspond to system 1, curves (2) to system 2 and curves (3) to system 3. We, also, present the line
corresponding to the HRG (curves (4)). It is seen that the HRG is the low temperature limit of systems 1 and 3,
something expected since at low temperatures the system exists in the form of hadrons.

As temperature rises the curves begin to deviate from the HRG curves. The total system reaches a maximum in 
$g_t$ and $\left<m\right>$ and then it begins to decrease. System 3 after reaching a local maximum, it 
decreases abruptly to arrive at a local minimum. Then it increases again to a local maximum and finally it stabilizes.
These variations of $g_t$ 
and $\left<m\right>$ occur in the temperature range of the chiral transition.
This behaviour is contrasted to a system with constant particle states. Such a system would have an effective
description with $g_t$ and $\left<m\right>$ always increasing when temperature increases. For example we can see 
the behaviour of the HRG in Figures \ref{fig:HRG} (ii) and (iii). The decrease of the number of states of 
systems 1 and 3 
suggests that the QCD system, as it is heated, it is driven to a gradual reduction of 
available states. The decrease of $\left<m\right>$, at certain temperature ranges, indicate that the average mass 
of these states decreases, also.
If the separation of the whole system into pure SU(3) sector and the remaining quark-gluon interacting system is approximately correct in the temperature range 154-230 MeV, then an important conclusion can be drawn. The degrees of freedom of the quark-gluon interacting system 3 drops to about 10. Then, this value cannot correspond to quark (coloured) degrees of freedom. The relevant value for the 3 flavours of quarks are 36. Even if we omit the heavy strange quark (on the assumption that its degrees of freedom will ``unlock'' at a greater temperature) the value for the 
remaining quarks cannot be less than 24 (in $2+1$ flavour QCD, $u$ and $d$ quarks are considered almost identical). As it is evident from Fig.~\ref{fig:gmx2}(i), in a model with two particle species, the effective number of states of the whole system can never be less than the number of
states of the lighter particle species. Consequently, the degrees of freedom of about 10, just above $T_c$, can only correspond to hadrons. 
So it is deduced that, at the critical temperature, the QCD system undergoes a drastic reduction of the hadronic degrees of freedom, possibly a reduction of the available resonances for the hadronic families.

System 2 exhibits a very steep increase as the critical temperature $T_c$ is approached from below.
Above $T_c$ the increase is turned to a very steep descend.

\begin{figure*}[h]
\centering
\ContinuedFloat
\includegraphics[scale=0.59,angle=0]{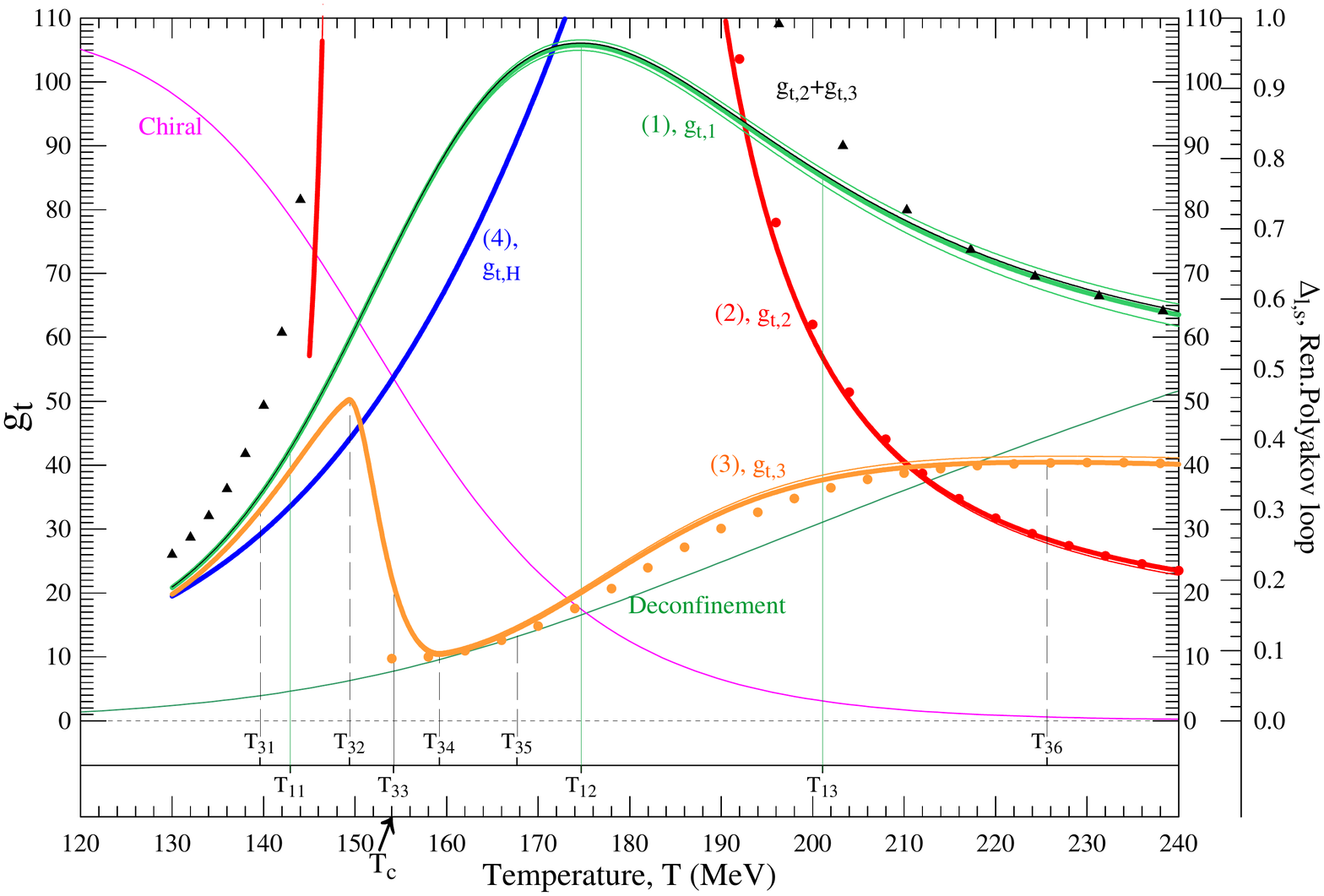}
\caption{\label{fig:Latgtfocus} {\small Fig.~\ref{fig:Latgt} focused on the temperature range 120-240 MeV.}}
\end{figure*}

\begin{figure*}[h]
\centering
\includegraphics[scale=0.59,angle=0]{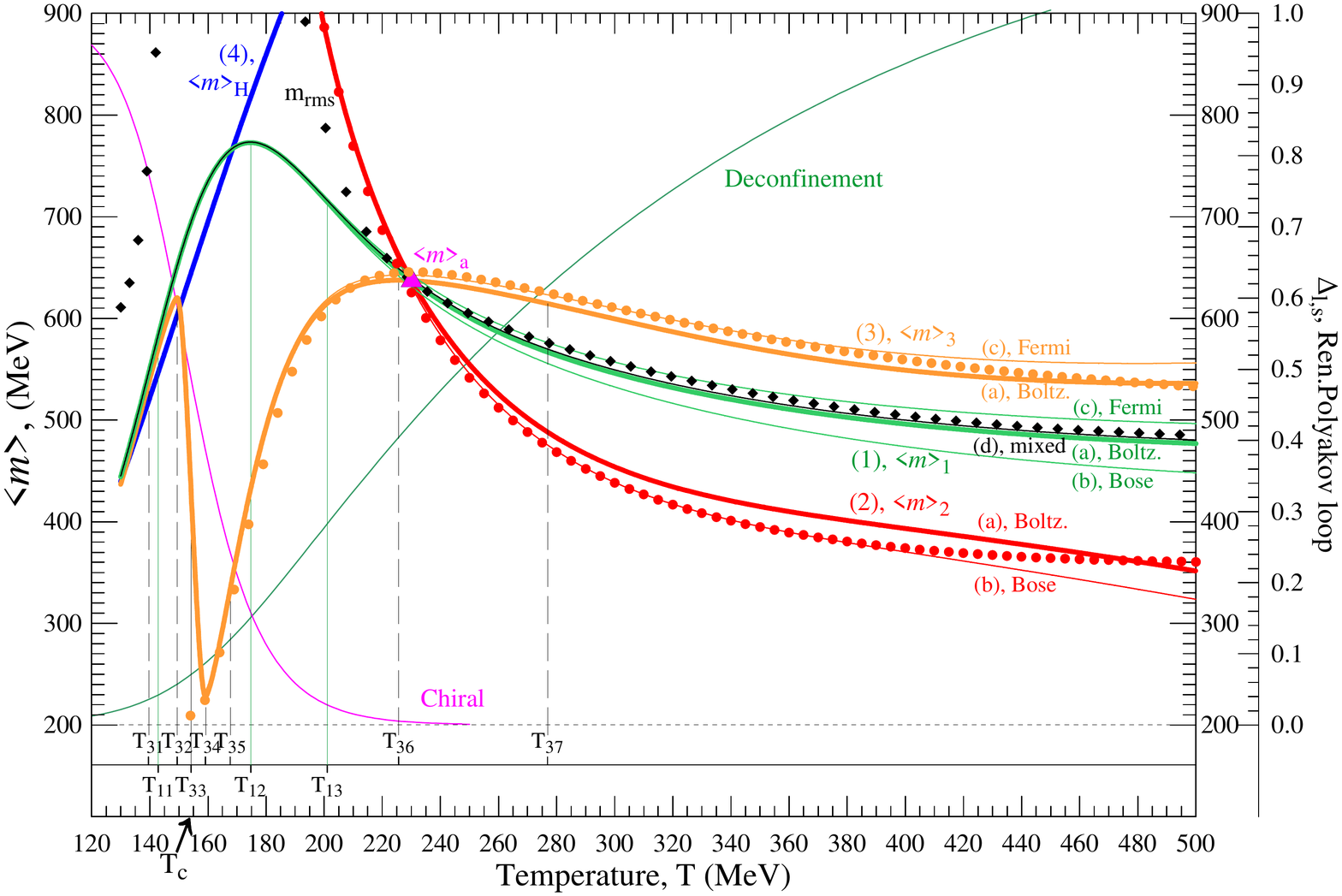}
\vspace{-0.5cm}
\caption{\label{fig:Latm} {\small The average mass $\left<m\right>$ calculated through method 1 with 
$\Delta T =5$ MeV. Curves (1), 
$\left<m\right>_1$,
correspond to
system 1, curves (2),
$\left<m\right>_2$, to
system 2, curves (3), 
$\left<m\right>_3$, to
system 3
and curve (4), $\left<m\right>_H$, to HRG. Thick lines (a) are calculated with the 
Boltzmann approximation, lines (b) with Bose statistics, lines (c) with Fermi statistics and thin
continuous 
line (d) to a system composed of $\frac{16}{52}$ bosons and of $\frac{36}{52}$ fermions. 
Solid rectangles represent the value of $m_{rms}$ calculated through the values of lines (2a) and (3a)
of this figure and figure \ref{fig:Latgt} and are compared with line (1a). 
Solid
triangle represents a point with common solution for average mass, $\left<m\right>_a$, for all
systems, 2, 3 and 1.
The order parameters of the chiral transition and deconfinement
are shown, as in Fig.~\ref{fig:P,s}. The shown characteristic temperatures 
$T_{31 \ldots 36}$,
$T_{11 \ldots 13}$
are defined in Fig.~\ref{fig:Latx2} and Table \ref{tab:x2_ex}.
With solid circles we present the calculations of $\left<m\right>$ in the Boltzmann approximation
for systems 2 and 3 carried out with the Lattice results from \cite{SU(3)_}.
}}
\end{figure*}

At temperatures greater than $\sim 250$ MeV, which is approximately the temperature where the chiral transition is 
completed, the number of states of all systems begins to approach an almost constant value. 
This suggests that the states of the system gradually cease to change. It is remarkable that the results of our fit 
reveals that the 
number states of system 2, $g_{t,2}$,
tends to the value $g_G=16$ of eq.~(\ref{eq:PgSB}) and the 
number states of system 3, $g_{t,3}$,
tends to the value $g_Q=36$ of eq.~(\ref{eq:PqSB}). As a result the total system also 
tends to the value $g_T=g_G+g_Q=52$ which is the total number of states of gluons and quarks. The values of $g_G$, $g_Q$
and $g_T$ are presented at the far right end of Figure \ref{fig:Latgt} as solid circles. Figure \ref{fig:Latm} shows 
that the average mass $\left<m\right>$ does not undergo great changes above $\sim 230$ MeV. The average mass of quarks,
even at $T=500$ MeV, remains high and is equal to $\simeq 540$ MeV. All these findings are consistent with a description
where, above $T \simeq 250$ MeV, colour states are present and their number is already locked to their corresponding 
high temperature limit. However, the system is still strongly interacting. This is indicated by the large value of
average mass, which still remains far from the zero value of the SB limit.

The curves (a) of Figs.~\ref{fig:Latgt} and \ref{fig:Latm} are calculated in the Boltzmann approximation (eq.~(\ref{eq:lnZfit_Bo})). 
To check the validity of these calculations we also perform calculations under the assumption that the system is composed 
of fermions or bosons. For this task we use as fitting partition function the following:
\begin{equation} \label{eq:lnZfit_BF}
\ln Z_{f,BF} (V,T; g_t,\left<m\right>,\alpha) = \frac{V}{6\pi^2 T}g_t \;\cdot
\int_0^{\infty} dp \frac{p^4}{\sqrt{p^2+\left<m\right>^2}} 
\left[\exp \left(\frac{\sqrt{p^2+\left<m\right>^2}}{T} \right)+ \alpha \right]^{-1} ,
\end{equation}

\begin{figure*}[h]
\centering
\ContinuedFloat
\includegraphics[scale=0.59,angle=0]{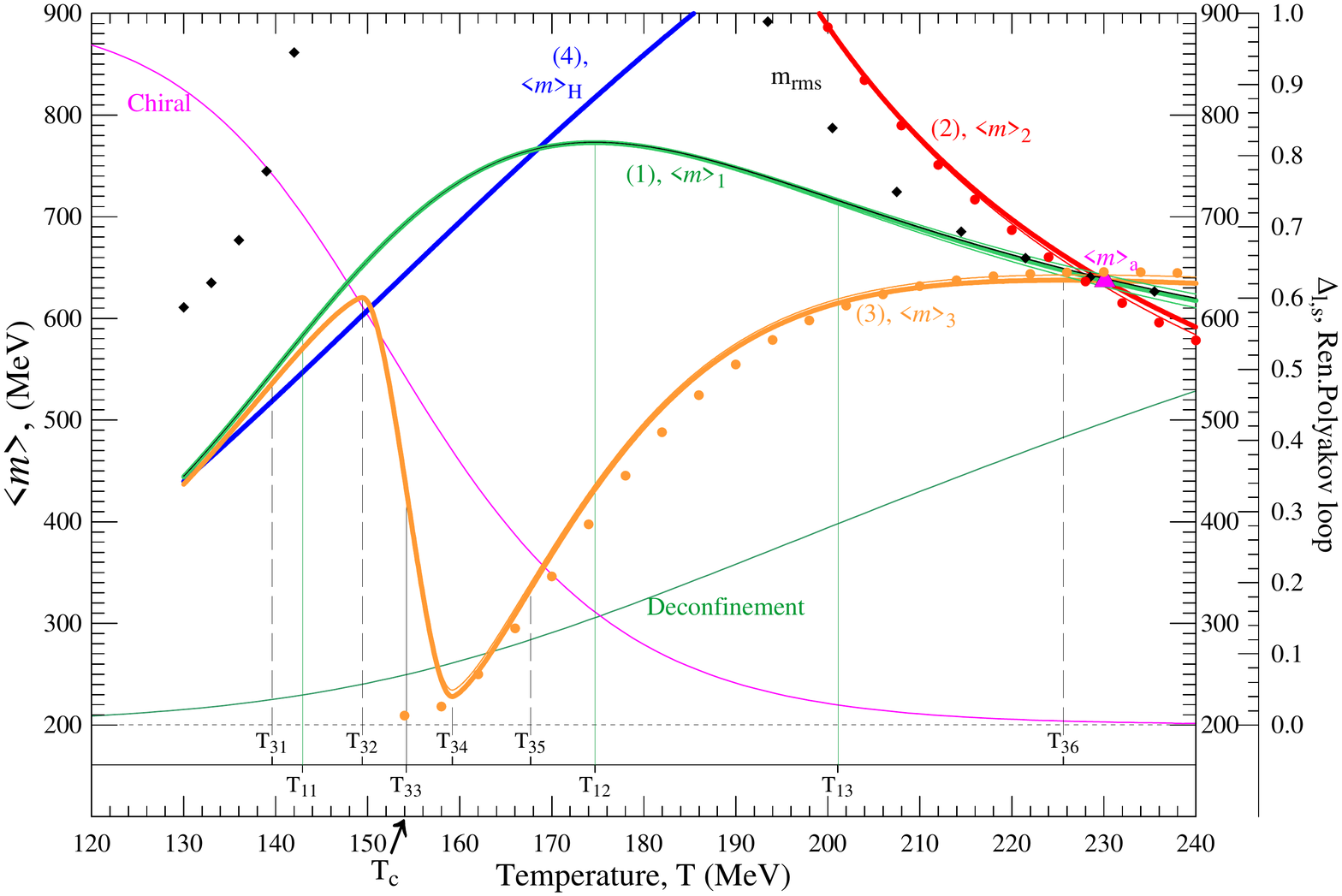}
\caption{\label{fig:Latmfocus} {\small Fig.~\ref{fig:Latm} focused on the temperature range 120-240 MeV.}}
\end{figure*}

\noindent where $\alpha=-1(+1)$ for bosons(fermions).
The results with the Bose statistics are represented by curves (b) and those with the Fermi statistics are represented
by curves (c). Since we know 
system 2
consists of bosons we present only a fit with the Bose 
statistics. For 
system 3
we know that it consists of fermions at high temperatures.
This system for low temperatures consists of hadrons, a mixture of fermions and bosons. However, at low 
temperatures all the statistics converge to the same results. For these reasons we present only a fit with the Fermi statistics for 
system 3.
For 
system 1
we present both statistics, as well as the result of a ``mixed'' fitting function: 
\begin{equation} \label{eq:lnZfit_m}
\ln Z_{f,m} (V,T; g_t,\left<m\right>) = \frac{16}{52} \ln Z_{f,BF} (V,T; g_t,\left<m\right>,-1)+
\frac{36}{52} \ln Z_{f,BF} (V,T; g_t,\left<m\right>,+1)
\end{equation}

All the results, with all types of statistics, are found to coincide for $T \leq 200$ MeV. At $T \simeq 500$ MeV 
the deviation between the Boltzmann approximation and the Bose/Fermi statistics is at most 8.6\%.
Also, the Fermi statistics for 
system 3
and the fitting function of eq.~(\ref{eq:lnZfit_m}) for 
system 1
drive the results of $g_t$ closer to the values $g_Q$ and $q_T$, respectively.

In Figure~\ref{fig:Latgt} we have also plotted the value of the sum $g_{t,g}+g_{t,q}$ for the Boltzmann 
calculations (solid triangles). This value almost coincides with the value for the total system 
$g_{t,1} \simeq g_{t,2}+g_{t,3}$, 
for temperatures $T>210$ MeV. The situation is similar to the simple system
of figure \ref{fig:gmx2}(i) (temperature is high enough so both particle species are present in the 
effective description). On the contrary, 
$g_{t,1}$
departs from the sum 
$g_{t,2}+g_{t,3}$
in the region
145 MeV$<T<$ 210 MeV, acquiring considerably less value than this sum. This is due to the high average mass of 
system 2.
Thus, in this region 
$g_{t,1}$
is closer to
$g_{t,3}$.

In Figure~\ref{fig:Latm} we present the calculation of $m_{rms}$ (solid rectangles) for a system which is 
composed of quarks and gluons:
\begin{equation} \label{eq:m_rms-gq}
m_{rms}=\sqrt{\frac{g_{t,2} \left<m\right>_2^2 + g_{t,3} \left<m\right>_3^2}{g_{t,2}+g_{t,3}}}\;.
\end{equation}
This value is very close to average mass of 
system 1 
$\left<m\right>_1 \simeq m_{rms}$ 
for temperatures $T>210$ MeV, situation similar to the simple system of figure \ref{fig:gmx2}(ii).

It is also 
seen
that at a certain temperature $T_a \simeq 230$ MeV all 
systems, 3, 2 and as a consequence 1,
obtain the same average mass $\left<m\right>_a \simeq 637$ MeV. This means that at 
this temperature all systems can be described with the same common effective mass. This mass appears
indicated with a solid triangle in Figure~\ref{fig:Latm}.

We found that at high temperatures the quark states approximately lock onto their value of degrees of freedom that
correspond to the free system, but with considerably high average mass. It would be interesting to seek the values
of masses which correspond to each of the two quark species ($u$ and $d$ quarks are considered identical and are 
denoted as $q$) at this high temperature limit. For
this reason we have performed a calculation at $T=$500 MeV. We use the pressure $P_3$ and entropy density $s_3$
at this temperature, according to eqs.~(\ref{eq:P3,s3}). We fix by hand the number of states for the $q$-quarks at 24 and for the $s$-quarks at 12. 
Subsequently we use the Fermi statistics and leave as parameters to be determined two different mass terms 
$\left<m\right>_q$ and $\left<m\right>_s$ for the two kinds of quarks. The result of the fit is
$\left<m\right>_q \simeq$482 MeV and $\left<m\right>_s \simeq$716 MeV, which correspond to 
$\left<m\right>_{rms} \simeq$571 MeV. The last value can be compared with the value one can find if only one mass 
parameter is used for all the quarks (with $g_{t3}=$36): $\left<m\right> \simeq$563 MeV (close to the
$\left<m\right>_{rms}$ value).

The calculated $\left<m\right>_q$ and $\left<m\right>_s$ are still well above the current quark masses  
$\sim$3.5 MeV and $\sim$98 MeV, as well as above the constituent quark masses $\sim$338 MeV and $\sim$486 MeV. 
This suggests that the quarks are still in interaction with the gluonic field at $T$=500 MeV.

Recently, new Lattice calculations for the pure SU(3) pressure and entropy density have been carried out \cite{SU(3)_}.
We have carried out the calculations for $g_t$ and $\left<m\right>$ for systems 2 and 3 with method

\begin{figure*}[h]
\centering
\includegraphics[scale=0.59,angle=0]{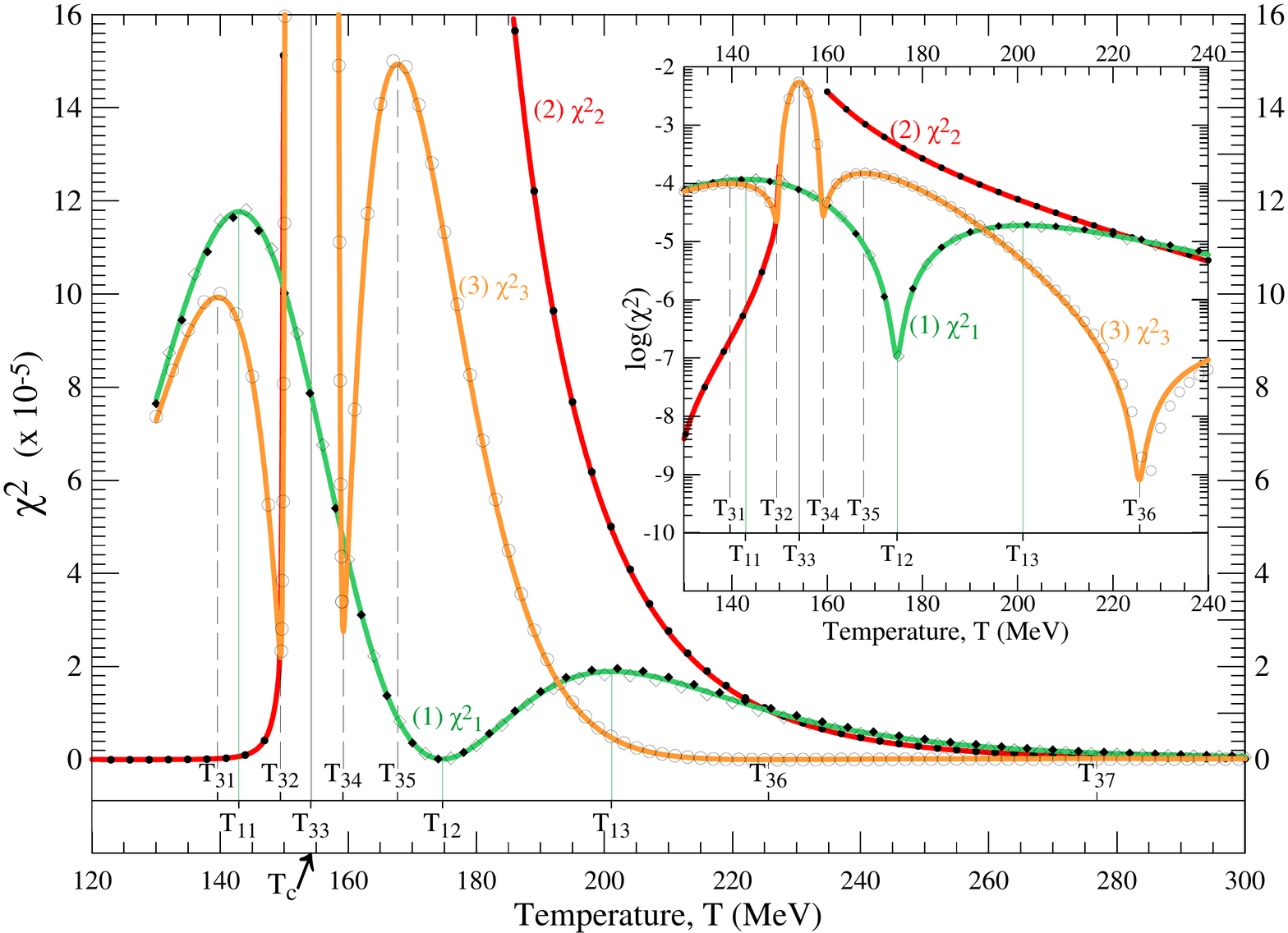}
\vspace{-0.2cm}
\caption{\label{fig:Latx2} {\small The value of $\chi^2$ which results from the fit to determine the
parameters $g_t$ and $\left<m\right>$ through method 1 with $\Delta T =5$ MeV in linear and logarithmic
vertical axis (embedded graph). 
Curve (1), 
$\chi^2_1$, corresponds to 
system 1, curve (2), 
$\chi^2_2$, to 
system 2 and curve (3),
$\chi^2_3$, to
system 3. 
Thick lines are calculated with the Boltzmann approximation, while solid dots represent the use of Bose 
statistics for
system 2 and open circles the use of Fermi statistics for
system 3. 
For 
system 1, temperatures
$T_{11}$ and $T_{13}$ represent location of maximum values of $\chi^2$ and
$T_{12}$ corresponds to minimum value.
For
system 3, temperatures
$T_{31}$, $T_{33}$, $T_{35}$ and $T_{37}$
represent location of maximum values of
$\chi^2$ and
$T_{32}$, $T_{34}$, $T_{36}$
correspond to minimum values. See also Table \ref{tab:x2_ex}.}}
\end{figure*}

\begin{table}[H]
\centering
\begin{tabular}{|c|c|c|c|c|} \hline
T     & Label    & System & Extremal    & Extremal \\
(MeV) &          &        & of $\chi^2$ & of $g_t$, $\left<m\right>$ \\ \hline
139.6 & $T_{31}$ & 3 & max & -   \\
142.9 & $T_{11}$ & 1 & max & -   \\
149.4 & $T_{32}$ & 3 & min & max \\
154.2 & $T_{33}$ & 3 & max & -   \\
159.2 & $T_{34}$ & 3 & min & min \\
167.7 & $T_{35}$ & 3 & max & -   \\
174.7 & $T_{12}$ & 1 & min & max \\
201.1 & $T_{13}$ & 1 & max & -   \\
225.6 & $T_{36}$ & 3 & min & min \\ 
276.9 & $T_{37}$ & 3 & max & -   \\ \hline
\end{tabular}
\caption{\label{tab:x2_ex} {\small Location of the extremal points of $\chi^2$ for 
systems 1 and 3,
as well as, their connection with the fitted parameters of Figures \ref{fig:Latgt} and \ref{fig:Latm}.
The maxima of $\chi^2$ depend on the choice of the errors $\sigma$, while the minima remain unchanged.
In this calculation we have used $\sigma=T^4$.}}
\end{table}

\begin{figure*}[h]
\centering
\includegraphics[scale=0.59,angle=0]{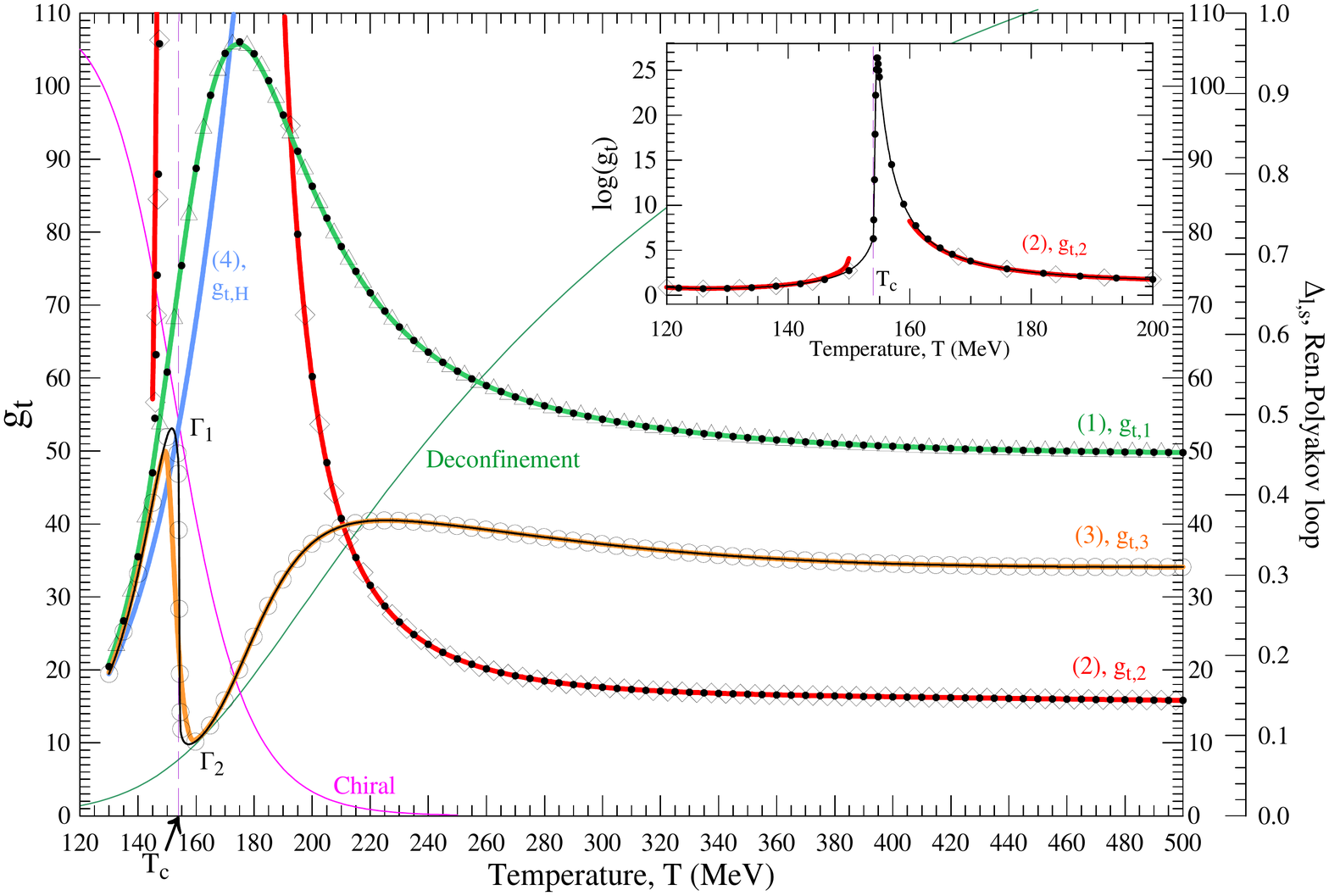}
\caption{\label{fig:Latgt_MDT} {\small The effect of the methodology of calculation on the number of states, $g_t$. 
All calculations in this graph have been carried out in the Boltzmann approximation.
Curves (1), 
$g_{t,1}$, correspond to
system 1, curves (2),
$g_{t,2}$, to 
system 2, curves (3),
$g_{t,3}$, to 
system 3 and curve (4), $g_{t,H}$, to HRG. 
Thick
continuous lines are calculated through method 1 with $\Delta T =5$ MeV.
Solid points near line (1) and (2) and solid
dark line near line (3) are the calculations through method 2 for 
system 1, 2 and 3, respectively.
Open symbols represent calculations through method 1 with $\Delta T =0.5$ MeV (triangles for
system 1,
rectangles for
system 2 and circles for
system 3).
The order parameters of the chiral transition and deconfinement are shown, as in Fig.~\ref{fig:P,s}.
Points $\Gamma_1$ and $\Gamma_2$ show regions of results sensitive to the methodology of calculation we used.
The embedded graph with the logarithmic vertical axis focuses on the calculations of
system 2
around critical temperature $T_c$.}}
\end{figure*}

\begin{figure*}[h]
\centering
\ContinuedFloat
\includegraphics[scale=0.59,angle=0]{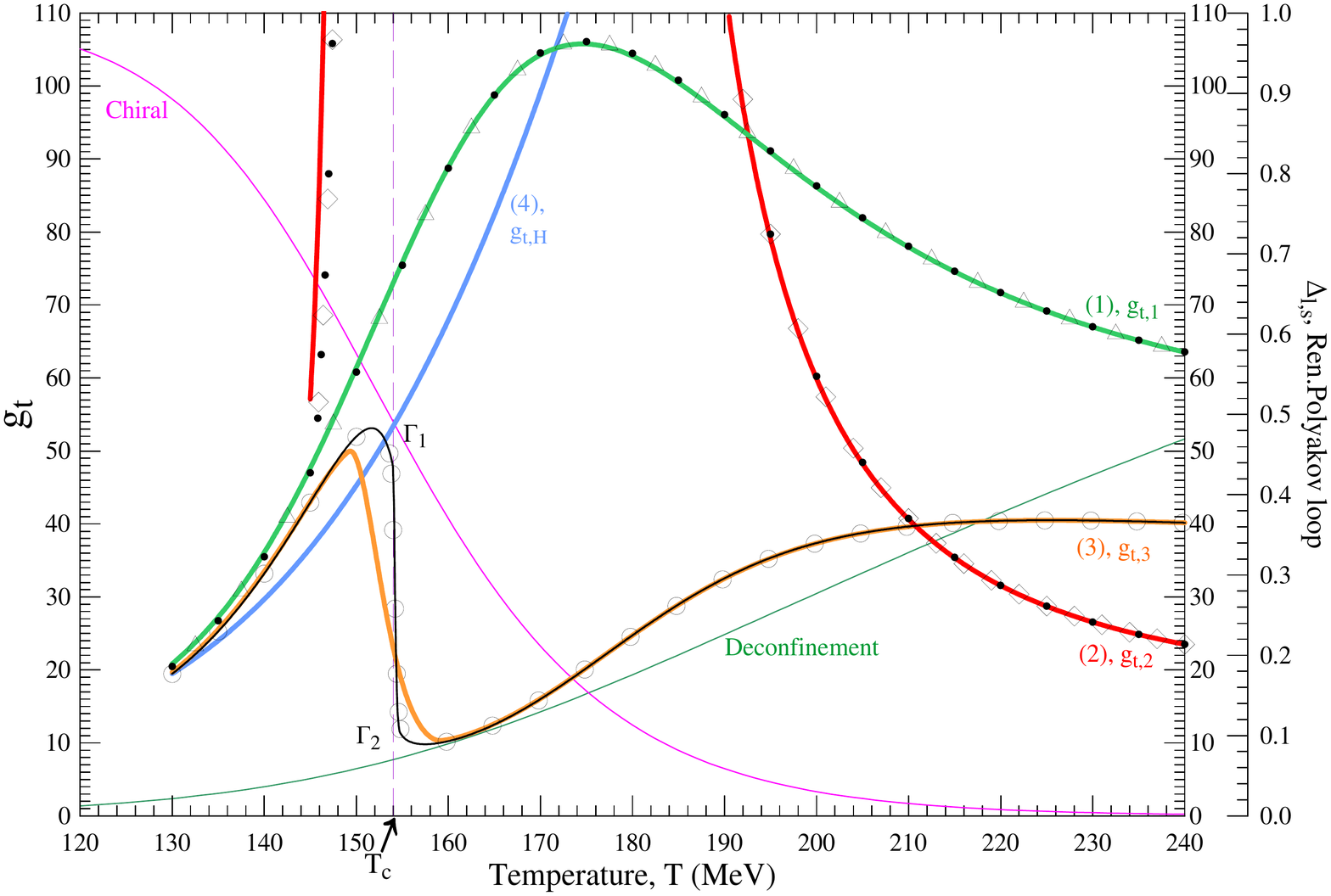}
\caption{\label{fig:Latgt_MDTfocus} {\small Fig.~\ref{fig:Latgt_MDT} focused on the temperature range 120-240 MeV.}}
\end{figure*}

\noindent
1 in the Boltzmann approximation,
which are depicted in Figs.~\ref{fig:Latgt} and \ref{fig:Latm}, respectively. 
The extracted parameters exhibit, apart from small variations with respect to the use of \cite{SU(3)}, the same overall behaviour. 
This finding contributes to the stability and consistency of our description.

In Figure~\ref{fig:Latx2} we examine the shape of $\chi^2$ as function of temperature for 
system 1
(curve (1)), 
system 2
(line (2)) and 
system 3
(line (3)). For the error $\sigma$ we have used
the function (\ref{eq:sigma}).
The temperatures where the local maximum values of $\chi^2$ are located, represent transition points from one effective description with $g_t$ and $\left<m\right>$ to another, as discussed in the method
section. 
The minimum values of $\chi^2$, on the contrary, correspond to conditions where the relatively best description is
achieved. It is found that the minimum values of $\chi^2$ coincide with local maxima or minima of $g_t$ 
and simultaneously $\left<m\right>$ with respect to temperature.
The minima of $\chi^2$ remain practically unchanged under different choices of the errors $\sigma$. We
have checked the validity of this fact with the choices $\sigma =const.$, $\sigma =T^4$ and 
$\sigma =P_L(T)$. The location of the maxima of $\chi^2$, however, depend on the choice of the function used for
the $\sigma$, but the existence of two successive stable minima will always lock the position of the corresponding
maximum between them.  
For 
system 1
two maximum values are found at 
$T_{11}$ and $T_{13}$
and one minimum at 
$T_{12}$.
For 
system 2,
method 1 fails to produce results in the region 150 MeV $<T<$ 160 MeV. However, as it is
evident from Figure~\ref{fig:Latx2}, an extreme maximum is formed in this region.
For
system 3
there are four maxima at 
$T_{31}$, $T_{33}$, $T_{35}$ and $T_{37}$. 
The highest maximum is located at 
$T_{33}$, 
which is close to the transition temperature $T_c$. There are also three minima, at 
$T_{32}$, $T_{34}$ and $T_{36}$.
The minimum at
$T_{36}$
is very close to temperature $T_a$ discussed above. The positions of the local 
extremal of $\chi^2$ for systems 1 and 3 are listed in Table \ref{tab:x2_ex}. These characteristic 
temperatures are, also, depicted in Figs.~\ref{fig:Latgt} and \ref{fig:Latm}.
From these two graphs it is deduced that the minimum values of $\chi^2$ coincide with local maxima or minima of $g_t$ and simultaneously $\left<m\right>$ with respect to temperature.

\begin{figure*}[h]
\centering
\includegraphics[scale=0.59,angle=0]{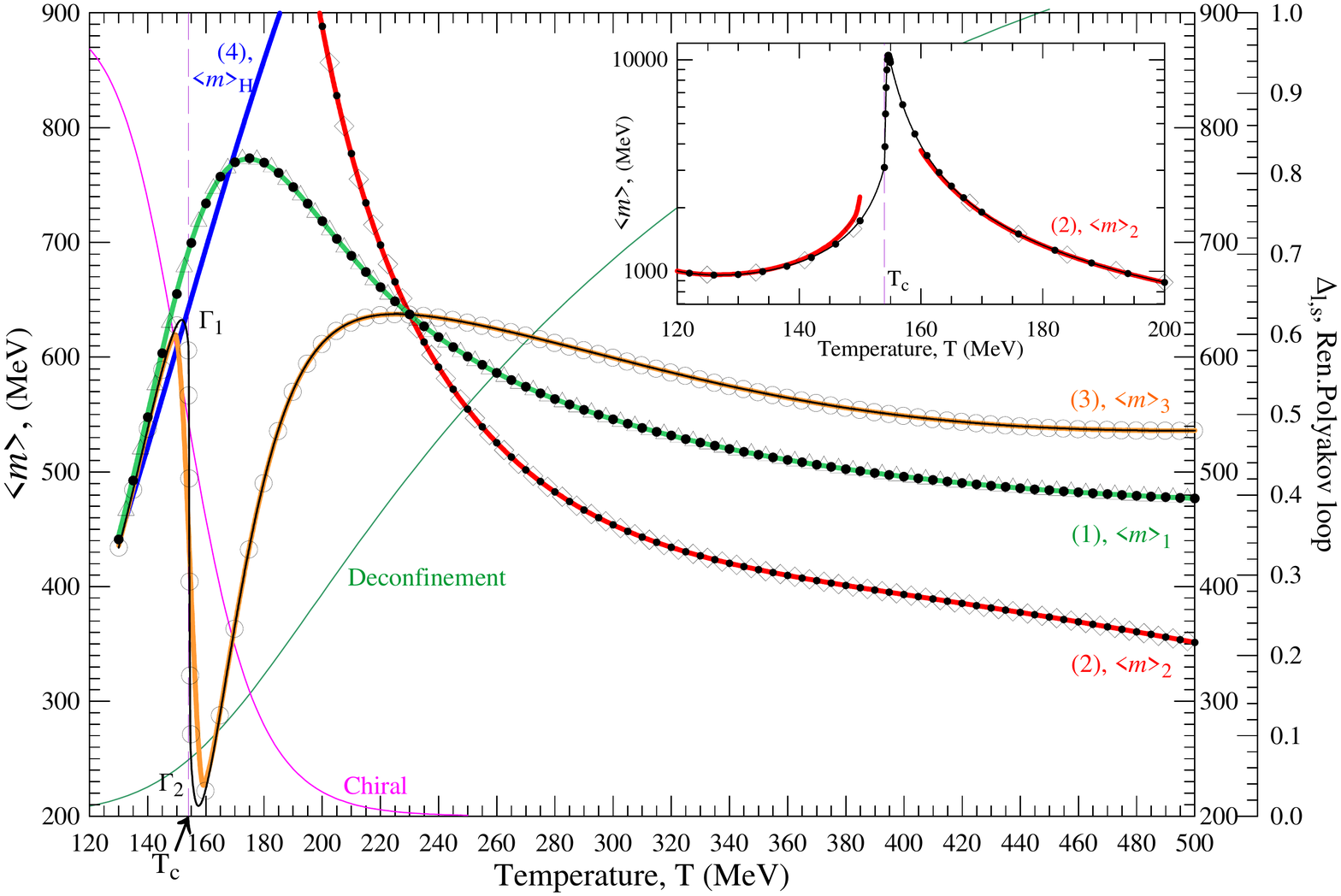}
\vspace{-0.5cm}
\caption{\label{fig:Latm_MDT} {\small The effect of the methodology of calculation on the average mass, $\left<m\right>$. 
All calculations in this graph have been carried out in the Boltzmann approximation.
Curves (1),
$\left<m\right>_1$, correspond to
system 1, curves (2),
$\left<m\right>_2$, to
system 2, curves (3),
$\left<m\right>_3$, to 
system 3 and curve (4), $\left<m\right>_H$, to HRG. 
Thick
continuous
lines are calculated through method 1 with $\Delta T =5$ MeV.
Solid points near line (1) and (2) and solid
dark
line near line (3) are the calculations through method 2 for
system 1, 2 and 3, respectively.
Open symbols represent calculations through method 1 with $\Delta T =0.5$ MeV (triangles for
system 1, rectangles for
system 2 and circles for
system 3).
The order parameters of the chiral transition and deconfinement are shown, as in Fig.~\ref{fig:P,s}.
Points $\Gamma_1$ and $\Gamma_2$ show regions of results sensitive to the used methodology of calculation.
The embedded graph with the logarithmic vertical axis focuses on the calculations of
system 2
around critical temperature $T_c$.}}
\end{figure*}

\begin{figure*}[h]
\centering
\ContinuedFloat
\includegraphics[scale=0.59,angle=0]{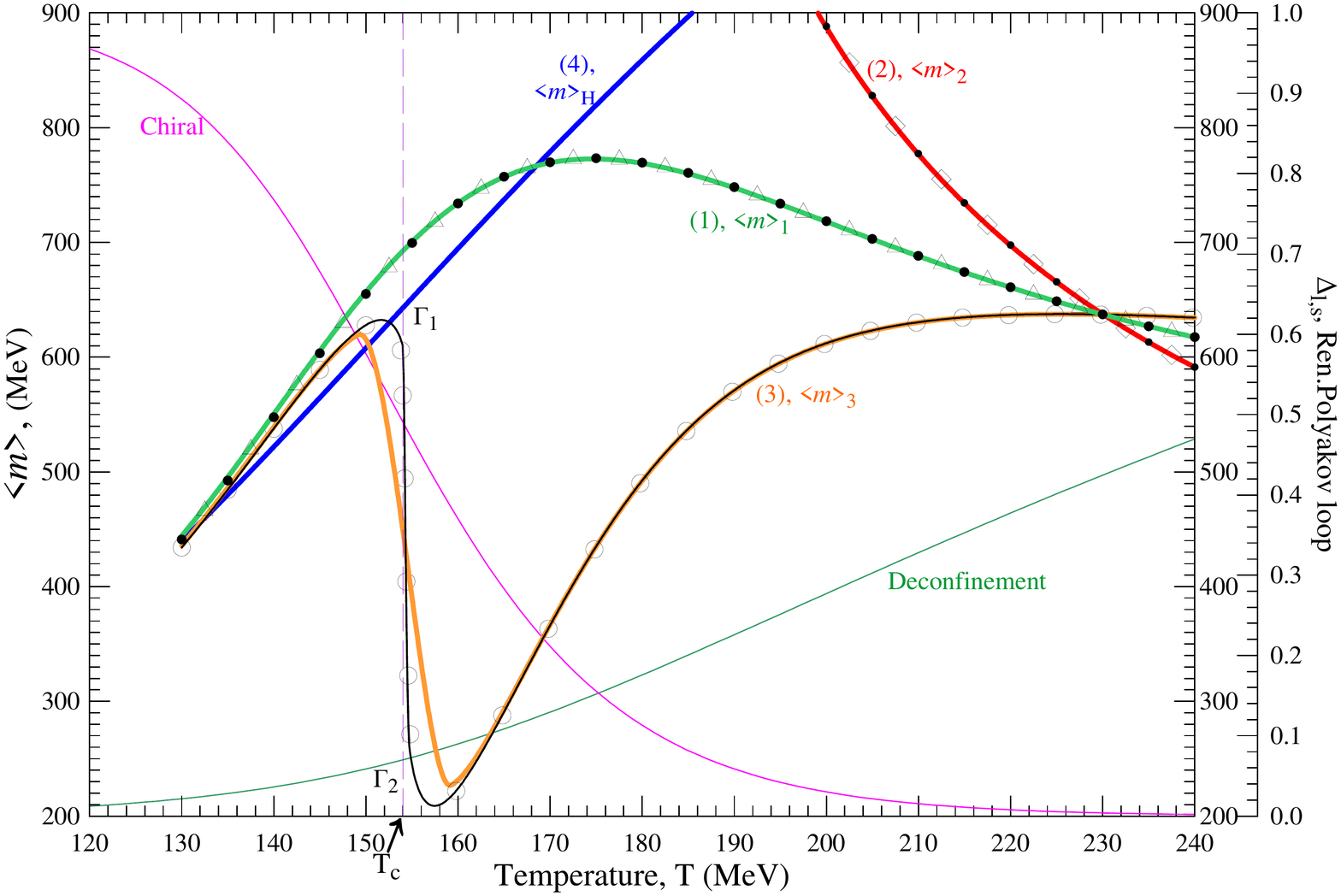}
\caption{\label{fig:Latm_MDTfocus} {\small Fig.~\ref{fig:Latm_MDT} focused on the temperature range 120-240 MeV.}}
\end{figure*}

\begin{figure*}[h]
\centering
\includegraphics[scale=0.59,angle=0]{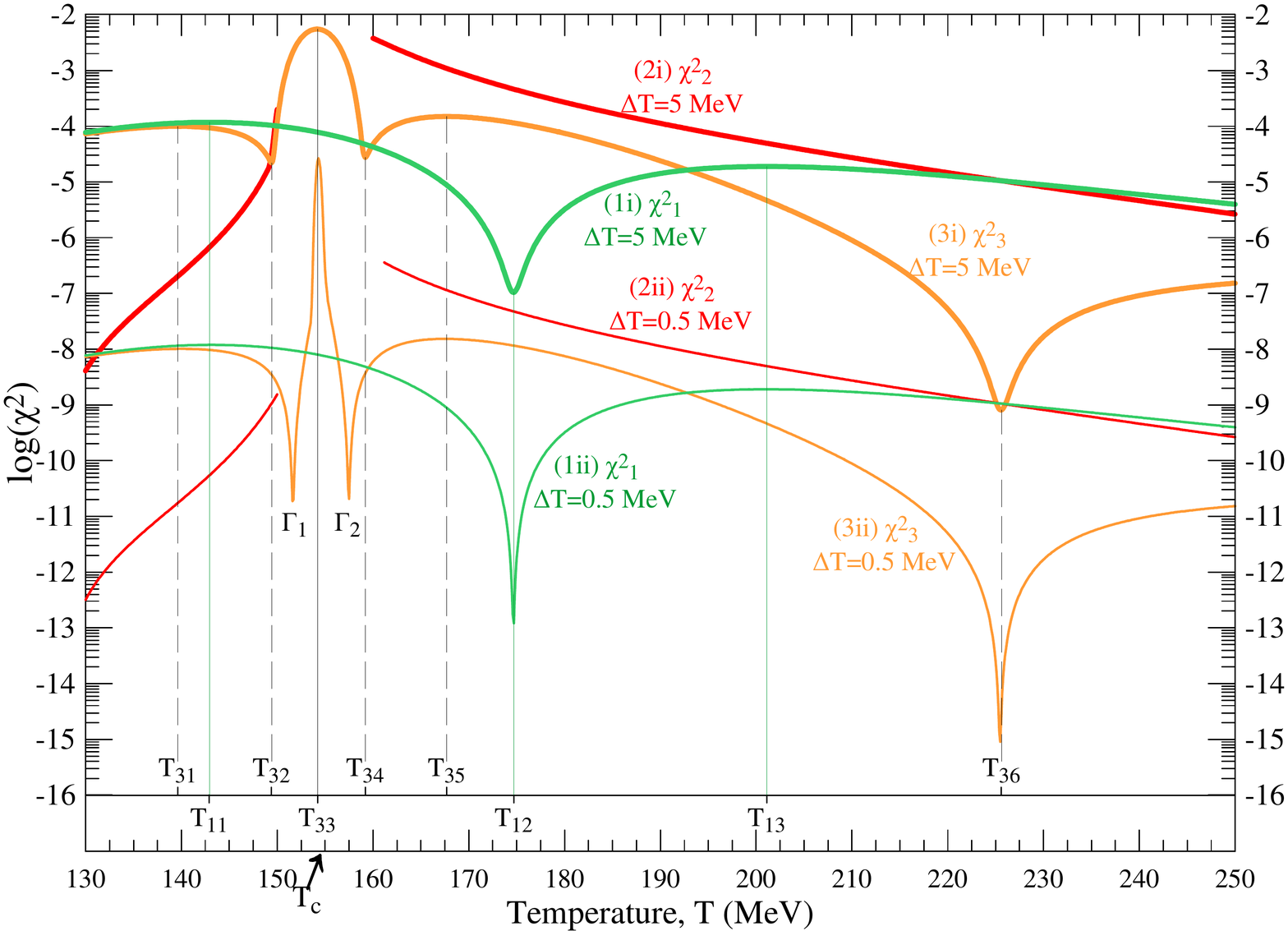}
\caption{\label{fig:Latx2_DT} {\small The effect of the methodology of calculation on the value of $\chi^2$.
The calculations have been carried out in the Boltzmann approximation.
Upper, thick curves (i) represent calculations through method 1 with $\Delta T =5$ MeV, while
low, thin curves (ii) represent calculations through method 1 with $\Delta T =0.5$ MeV.
Curves (1),
$\chi^2_1$, correspond to
system 1, curves (2), 
$\chi^2_2$, to
system 2 and curves (3), 
$\chi^2_3$, to
system 3. 
The shown characteristic temperatures are the same as in Figure~\ref{fig:Latx2} and are all stable under change of
$\Delta T$, except for
$T_{32}$ and
$T_{34}$.}}
\end{figure*}

The value of $\chi^2$ can also reveal the quality of the fit. In order to estimate the value of $\chi^2$ to be 
used for this purpose we have to use for the errors $\sigma$ in eq.~\ref{eq:chi2} a value comparable to the sum
$\sum_{i=1}^{N}\left[P_L(T_i)-P_f(T_i)\right]^2$. One suitable choice is $\sigma'=0.1 P_L(T)$, where $T$ is the central
temperature in the aforementioned sum and the value $0.1$ corresponds to a typical relative experimental error.
The value of $\chi^2$ also needs to be divided to the corresponding degrees of freedom ($dof$) which are 99
(101 data points minus 2 fitted parameters).
The quality of the fit may then be estimated by the quantity
\begin{equation} \label{eq:chi2norm}
\frac{\chi'^2}{dof}=\frac{\chi^2}{dof} \left[\frac{\sigma}{0.1 P_L(T)}\right]^2\;.
\end{equation}
If we exclude 
system 2
around the critical temperature $T_c$, it is seen that the highest value of 
$\chi^2$ for all
systems
corresponds to 
system 3
near $T_c$ (see fig.~\ref{fig:Latx2}).
For this value of $\chi^2=5.51 \cdot 10^{-3}$ we calculate $\frac{\chi'^2}{dof}=1.15 \cdot 10^{-4}$. The
value of $\frac{\chi'^2}{dof}$ is considerably lower for the rest of the temperatures. Thus, we conclude that we have
a successful effective description for all
systems
in all temperatures, with the exception of
system 2
in the vicinity of $T_c$. 

We investigate further the stability of  $\chi^2$ under the influence of change of the statistics. In 
Figure~\ref{fig:Latx2} we have plotted the evaluation of $\chi^2$ for
system 2
with the use of Bose statistics (solid points) and for 
system 3
with Fermi statistics (empty circles). No apparent
change is found (with the exception of the small shift of the minimum 
$T_{36}$),
proving that $\chi^2$ remains practically unaltered by the change of the type of statistics used.

In the following we inspect the influence on our results of changing $\Delta T$ in method 1 (the
temperature interval around the central temperature $T$ on which we impose our fits), as well as using
method 2, instead of method 1. In Figures \ref{fig:Latgt_MDT} and \ref{fig:Latm_MDT} we show results for
$g_t$ and $\left<m\right>$, respectively. We have plotted by thick 
continuous
lines the results of method 1 with $\Delta T$=5 MeV. The results of method 2 are plotted by solid points for 
systems 1 and 2
and by thin 
continuous
line for 
system 3.
The results of method 1 with $\Delta T$
reduced to 0.5 MeV are shown with open symbols, triangles for 
system 1,
rectangles for 
system 2
and circles for
system 3.
We find complete stability of our results in the whole temperature range of our 
analysis under changing of the method or the magnitude of $\Delta T$. The only exception appears in the
abrupt variation of $g_t$ and $\left<m\right>$ in the case of
system 3
(see points $\Gamma_1$ and
$\Gamma_2$) around the critical temperature $T_c$. The reduction of the temperature interval $\Delta T$ or 
the use of method 2 leads to a steepest descend of our fitted parameters with temperature increase. The
equivalence of method 2 with the reduction of $\Delta T$ is expected, since method 2 is identical to
method 1 for $\Delta T$=0.

In Figures \ref{fig:Latgt_MDT} and \ref{fig:Latm_MDT} we have embedded a figure to show the behaviour of
system 2
in the temperature interval (150-160) MeV. Method 1 fails at this region, but method 2 is able
to produce results, which are shown in the graph. The extreme rise of the parameters $g_t$ and $\left<m\right>$
occurs around the critical temperature $T_c$. The exact values of the parameters are not trustworthy in this 
region. The reason is that we have used the smoothing interpolating function \ref{eq:fitsu3middle} and that
$\chi^2$ of method 1 is forming an extreme maximum as can be seen in Fig.~\ref{fig:Latx2}.
However, it indicates qualitatively an extreme increase of the number of states and the effective mass of
the gluons around the critical temperature. 
The emerging picture is as if the gauge field occupies the whole system as a unique entity.
This system, around the critical temperature, cannot be described as a group of ``particles'' with specific
mass and states, since the interaction is such that the whole system behaves as a single ``particle'' with
divergent mass and number of states.

Finally, in Figure \ref{fig:Latx2_DT} we check the effect on $\chi^2$ of changing $\Delta T$ in method 1.
In this Figure we present the calculations for $\Delta T=5$ MeV (lines (i), upper) and $\Delta T=0.5$ MeV 
(lines (ii), low). We find that $\chi^2$ is decreasing as $\Delta T$ reduces. This is expected, since in
method 2 we have identically $\chi^2=0$ (two parameters are determined through solving two equations) and
also $\Delta T=0$. Thus, we expect that the reduction of $\Delta T$ causes $\chi^2$ to decrease, since both
of these non-negative quantities should approach the zero limit.
The figure shows that the extremal points of $\chi^2$ remain unchanged under
variation of $\Delta T$. The only exception appears to the minima 
$T_{32}$ and $T_{34}$ of system 3.
Reduction of $\Delta T$ causes both of them to approach the location 
$T_{33}$
of the corresponding maximum
of $\chi^2$. The behaviour is consistent with figures \ref{fig:Latgt_MDT} and \ref{fig:Latm_MDT}, where
the reduction of $\Delta T$ leads to steepest variation of $g_t$ and $\left<m\right>$ around $T_c$.
Consequently, this leads to narrowing the gap between the location of maximum and minimum values of these 
parameters.

\section{Concluding Remarks}

We have developed an effective description of the Lattice QCD pressure and specific entropy at zero 
baryon chemical potential with two parameters, the degeneracy factor, $g_t$ and the average particle mass 
$\left<m\right>$. The description is carried out for the
three flavour QCD system (system 1),
as well as the
pure gauge field (system 2)
 and
a system containing quarks in interaction with gluons (system 3).

The calculated parameters of 
systems 1 and 3
have as their low temperature limit the 
corresponding parameters of the Hadron Resonance Gas. The number of states for all 
systems (1, 2 and 3)
converge, above $T \simeq 230$ MeV, close to the number of states of an ideal Quark-Gluon
Phase, indicating the existence of colour states at these temperatures. The corresponding high average masses, 
though, suggest that the entities are strongly interacting.

The number of states and the average mass, corresponding to
system 2,
are found to be in extreme
increase just below the critical temperature, $T_c$ and in extreme decrease just above $T_c$.

Between $T \simeq 150$ and 160 MeV the average mass and the number of states 
corresponding to system 3
are found to decrease steeply, deviating from the HRG parameters. 
However, the minimum value of degrees of freedom at this temperature interval indicates the
existence of hadronic states. 
One possible explanation of this behaviour would be
the reduction of the available hadron resonances and their limitation to a few states. Between 160 MeV and
$\sim220$ MeV the number of states is increasing approaching an almost constant value, equal to that
of a system of three flavour, coloured quark states.
The described changes of 
systems 3 and 2
are taking place within the temperature range of the chiral transition.

The value of $\chi^2$ which results after the optimization of the parameters $g_t$ and $\left<m\right>$
can be used as a tool to locate regions of relatively best effective descriptions in terms of these parameters,
as well as points of transition between them. 
The low values of this quantity, after proper normalization, ascertains the
success of the description with the fitted parameters.

\vspace{0.5cm}
{\bf Acknowledgement.} We wish to thank N.~G.~Antoniou and F.~K.~Diakonos for useful suggestions and
fruitful discussions.


\begin{thebibliography}{}

\bibitem{Hagedorn} R.~Hagedorn,``Thermodynamics of Strong Interactions'', Lectures Given  in the Academic Training Programme 
of CERN 1970-71, CERN 71-12, 7 May 1971

\bibitem{quasi-particle} E.P.~Politis, C.E.~Tsagkarakis, F.K.~Diakonos, X.N.~Maintas and A.~Tsapalis, Phys.~Lett.~B {\bf 763} 
(2016) 139.

\bibitem{2+1flavor} A.~Bazavov et.~al., (HotQCD Collaboration), Phys.~Rev.~D {\bf 90} (2014) 094503;
arXiv:1407.6387 [hep-lat].

\bibitem{HRG} J.~Cleymans, H.~Satz, Z.~Phys.~C {\bf 57} (1993) 135;
J.~Letessier, A.~Tounsi, J.~Rafelski, Phys.~Lett.~B {\bf 292} (1992) 417;
J.~Letessier, A.~Tounsi, U.~Heinz, J.~Sollfrank, J.~Rafelski, Phys.~Rev.~D {\bf 51} (1995) 3408;
F.~Becattini, M.~Ga\'{z}dzicki, A.~Ker$\ddot{\rm a}$nen, J.~Manninen, R.~Stock, Phys.~Rev.~C {\bf 69} (2004) 024905;
J.~Cleymans, K.~Redlich, Phys.~Rev.~Lett.~{\bf 81} (1998) 5284.

\bibitem{SU(3)} Sz.~Bors\'{a}nyi, G.~Endr\H{o}di, Z.~Fodor, S.D.~Katz and K.K.~Szab\'{o}, JHEP {\bf 1207} (2012) 056; 
arXiv:1204.6184 [hep-lat].

\bibitem{ch-d} Sz.~Bors\'{a}nyi, Z.~Fodor, C.~Hoelbling, S.D.~Katz, S.~Krieg, C.~Ratti and K.K.~Szab\'{o}, JHEP {\bf 2010} (2010) 73;
arXiv:1005.3508 [hep-lat].

\bibitem{SU(3)_} L.~Giusti and M.~Pepec, Phys.~Lett.~B {\bf 769} (2017) 365; arXiV:1612.00265 [hep-lat].

\end{thebibliography}
\end{document}